\documentclass[12pt]{article}
\usepackage{caption,amsmath,graphicx,url,amssymb,latexsym,epsfig,tabularx,setspace,multirow,threeparttable,longtable,pdflscape,tabu,appendix,comment,todonotes,subfigure} 
\usepackage{algpseudocode}
\usepackage{algorithm}  
\usepackage{color,soul}

\usepackage[margin=1in]{geometry}
\usepackage{xspace}

\algnewcommand\algorithmicabort{\textbf{abort}}
\algnewcommand\Abort{\algorithmicabort}

\begin{document}
\def\gpuFSG{\textsc{SearchSpatial}\xspace}
\def\gpuS{\textsc{GPUSpatial}\xspace}

\def\gputemporal{\textsc{SearchTemporal}\xspace}
\def\gpuT{\textsc{GPUTemporal}\xspace}

\def\gpuspatiotemporal{\textsc{SearchSpatiotemporal}\xspace}
\def\gpuST{\textsc{GPUSpatioTemporal}\xspace}

\def\random{\textit{Random-1M}\xspace}
\def\merger{\textit{Merger}\xspace}
\def\dense{\textit{Random-dense}\xspace}

\begin{centering}
\textbf{Towards Efficient Indexing of Spatiotemporal Trajectories on the GPU for Distance Threshold Similarity Searches\\}
\vspace{0.3cm}
Michael Gowanlock\\ 
Department of Information and Computer Sciences and NASA Astrobiology Institute\\ University of Hawai`i, Honolulu, HI, U.S.A.\\
Email: gowanloc@hawaii.edu\\
\vspace{0.3cm}
Henri Casanova\\
Department of Information and Computer Sciences\\ University of Hawai`i, Honolulu, HI, U.S.A.\\
Email: henric@hawaii.edu

\end{centering}

\begin{abstract}
Applications in many domains require processing moving object trajectories.  In this work, we focus on a trajectory similarity search that finds all trajectories within a given distance of a query trajectory over a time interval, which we call the distance threshold similarity search.  We develop three indexing strategies with spatial, temporal and spatiotemporal selectivity for the GPU that differ significantly from indexes suitable for the CPU, and show the conditions under which each index achieves good performance.  Furthermore, we show that the GPU implementations outperform multithreaded CPU implementations in a range of experimental scenarios, making the GPU an attractive technology for processing moving object trajectories.  We test our implementations on two synthetic and one real-world dataset of a galaxy merger.           

\end{abstract}

\section{Introduction}
Trajectory data is generated in a wide range of application domains, such as the motions of people or objects captured by global positioning systems (GPS), the movement of objects in scientific applications, such as stars in astrophysical simulations, vehicles in traffic studies, animals in zoological studies and a range of applications of geographical information systems (GIS).  We study historical continuous trajectories \cite{Forlizzi2000}, where a database of trajectories is given as input and is searched to gain domain-specific insight.  In particular,  we study the \emph{distance threshold search}: Find all trajectories within a distance $d$ of a given query trajectory over a time interval [$t_{start}$,$t_{end}$].  An example of this search would be to find all prey within 200 m of all predators over the period of a month.

The challenges associated with moving object trajectories in comparison to stationary objects has prompted a literature on efficient trajectory indexing and processing strategies.  Many of the methods developed by the spatial and spatiotemporal database communities focus on sequential implementations, where a fraction of the data is stored in memory, and the rest is stored on disk.  Thus, reduction of disk accesses is the main optimization goal in these works. Alternatively, with relatively large memories available in modern workstations, sizable in-memory databases have become feasible.  Furthermore, with the proliferation of multicore and manycore architectures, parallel in-memory implementations can provide significant performance improvements over sequential out-of-core implementations. In instances where memory capacity on a single node is insufficient, historical continuous trajectory datasets can be partitioned and queried in-memory across multiple compute nodes in parallel.

To this end, we study the efficient processing of distance threshold
searches on trajectory databases using General Purpose Computing on Graphics
Processing Units (GPGPU). We focus on developing and comparing the
performance of GPU-friendly indexing strategies, and make the following
contributions:

\begin{itemize}
\item We develop three indexing techniques that are suitable for distances threshold searches on the GPU.
\item For each of the indexes, we develop an associated GPU kernel that minimizes branch instructions to achieve good parallel efficiency.
\item We compare our GPU implementation to a previously developed CPU-only implementation that uses an in-memory R-tree index, and show that using the GPU can afford significant speedup.

\item We find that when using large datasets, in contrast to smaller
datasets previously used in the literature, efficient trajectory splitting
strategies for an R-tree index, at least for the in-memory case, provides
limited or no performance improvements.

\item We evaluate our algorithms and kernel implementations  with 4-D datasets (3 spatial dimensions and 1 temporal dimension), including a real-world astrophysics dataset (of a galaxy merger) and two synthetic datasets.
\end{itemize}     

The paper is outlined as follows: Section~\ref{sec:related_work} outlines a motivating example and discusses related work.  Section~\ref{sec:problem_def} formally defines the problem. Section~\ref{sec:index} describes our three indexing techniques and search algorithms.  Section~\ref{sec:exp_eval} presents our experimental results. Finally, Section~\ref{sec:conclusions} concludes with a summary of our findings and a discussion of future research directions.

\section{Background and Motivating Example}\label{sec:related_work}
\subsection{Motivating Example}
One motivating application for this work is in the area of
astrophysics/astrobiology~\cite{2011AsBio..11..855G}. Astrobiology is the
study the evolution, distribution and future of life in the universe.
Biologists study the habitability of the Earth and find that life can
exist in a multitude of environments (including extreme environments, such
as temperature, pressure, salinity, radiation exposure, and others).  The
past decade of exoplanet searches implies that the Milky Way, and hence the
universe, hosts many rocky, low mass planets that may be capable of
supporting complex life (land-based animal life).   Given that there are
many planets in the Milky Way and given the broad range of conditions in
which life is found to thrive on Earth, the notion of the Galactic
Habitable Zone has emerged, i.e., the region(s) of the Galaxy that may
favor the development of complex life.  With regards to long-term
habitability, some regions of the Milky Way may be inhospitable due to
transient radiation events, such as supernovae explosions or close
encounters with flyby stars that can gravitationally perturb planetary
systems. Studying habitability thus entails solving the following two types
of \emph{distance threshold searches} on the trajectories of (possibly
billions of) stars orbiting the Milky Way: (i)~Find all stars within a
distance $d$ of a supernova explosion (or gamma ray burst), i.e., a
non-moving point over a time interval; and  (ii)~Find the stars, and
corresponding time periods, that host a habitable planet and are within a
distance $d$ of all other stellar trajectories.

\subsection{Background and Related Work}

A key question in database research is the efficient retrieval of data. In
the most general context, database management systems provide information
about database content and support arbitrary queries. However, in specific
domains it is possible to achieve more efficient retrieval if there are
structures and constraints on the data stored in the database and/or if
particular types of queries are expected.  Such a domain is that of spatial
and spatiotemporal databases that store the trajectories of moving objects.
A trajectory is a collection of points associated with the positions of an
object over time, where the points are connected by polylines (line
segments).  Such data, which arises in many scientific domains but is also
pervasive in modern society (GPS data, GIS applications), presents both
opportunities and challenges that are studied in the spatiotemporal
database community. The main goal of these databases is to perform
\emph{trajectory similarity searches}, i.e., finding trajectories within a
database that exhibit similarity in terms of spatial and/or temporal
proximity, or exhibit similarity in terms of spatial and/or temporal
features so that trajectories can be classified as belonging to a certain
group.   Similarity searches have been studied in various domains,
such as convoys \cite{Jeung:2008:DCT:1453856.1453971}, flocks
\cite{Vieira:2009:ODF:1653771.1653812}, and swarms
\cite{Li:2010:MMM:1807167.1807319}.  A predominant trajectory similarity
search that is used in many application areas is the $k$NN ($k$ Nearest
Neighbors) search~\cite{Frentzos2005,Frentzos2007,Gao2007,Guting2010}.

The typical approach in previous spatiotemporal database works proceeds in
two phases: (i)~search an index to obtain a preliminary result set;
(ii)~use refinement to produce the final result set. The search phase
focuses on \emph{pruning}, i.e., avoiding parts of the index based on the
selecting criteria of the query. To this end, several index-trees have been
proposed inspired by the success of the popular
R-tree~\cite{Guttman-R_tree}, such as TB-trees~\cite{Pfoser2000},
STR-trees~\cite{Pfoser2000}, 3DR-trees~\cite{Theodoridis-1996},
SETI~\cite{Chakka2003}, and implemented in systems such as
TrajStore~\cite{Cudre-Mauroux2010} and SECONDO~\cite{Guting2010}.  More
specifically, these works map nodes in an index-tree to pages stored on
disk. Performance is a function of the number of index-tree nodes that are
accessed, aiming to keep this number low so as to avoid avoiding costly
data transfers between memory and disk. Index-trees have been used
extensively for $k$NN searches.

In this work we study distance threshold searches, which can be viewed as
$k$NN searches with an unknown value of $k$ and thus unknown result set
size. As a result, several of the aforementioned index-trees, while
efficient for $k$NN searches, are  not efficient for distance threshold
searches. This is because, as $k$ is unbounded, standard index pruning methods cannot be
used.  Distance-threshold searches, although relevant to several application
domains, have not received a lot of attention in the
literature.  Our previous work in~\cite{Gowanlock2014} studies in-memory
sequential distance threshold searches, using an R-tree to index
trajectories inside hyperrectangular minimum bounding boxes (MBBs). The main
contribution therein is an indexing method that achieves a desirable trade-off
between the index overlap, the number of entries in the index, and the
overhead of processing candidate trajectory segments.
The work in~\cite{Arumugam2006} solves a similar problem, i.e., finding
trajectories in a database that are within a query distance $d$ of a search
trajectory and the authors propose four query processing strategies. A key
difference with the work in~\cite{Gowanlock2014} is that part of the
database resides on disk.  Other trajectory similarity
searches rely on metrics of similarity at coarse grained
resolutions ~\cite{Giannotti:2007}. 
Instead, the similarity search we study in this work necessitates precise
comparisons between individual polylines, to find the exact time
intervals when trajectories are within the threshold distance.
The large number of such comparisons is a motivation for using the GPU.

In the context of in-memory moving object trajectory databases several authors have
explored the use of multicore and manycore architectures.  Spatial and
spatiotemporal indexing methods have been advanced for use on the
GPU~\cite{Zhang2014,Zhang:2012:USH:2390226.2390229,You:2013:PSQ:2534921.2534949,Luo2012}.
Given the single instruction multiple data (SIMD) nature of the GPU, proposed
indexes for this architecture tend to be less sophisticated than the index-trees
used in the context of out-of-core databases.  This is in part because
branches in the instruction flow cause thread serialization and thus
loss of parallel efficiency~\cite{Han:2011:RBD:1964179.1964184}.  The $k$NN query
(not on trajectories) has been studied in the context of the
GPU~\cite{Pan:2011:FGL:2093973.2094002,CPE:CPE1718} and on hybrid CPU-GPU
environments~\cite{Krulis2012}.  In this work we focus on indexing techniques for
distance threshold similarity searches on trajectories for the GPU, which
to our knowledge has only been explored in our previous
work~\cite{Gowanlock2014c}.  
That previous work focuses on a scenario in which the query set cannot fit
entirely on the GPU due to memory constraints, thereby requiring
back-and-forth communication between the host and GPU, and thus a
particular indexing scheme.  Instead, in this work,  the query set fits on the
GPU, which makes it possible to explore a range of indexing schemes (while still having to consider memory constraints).

\section{Problem Statement}\label{sec:problem_def}

\subsection{Problem Definition}

Let $D$ be a spatiotemporal database that contains $n$ 4-dimensional (3
spatial and 1 temporal dimensions) \emph{entry line segments}. A
line segment $l_i$, $i=1,\ldots,|D|$, is defined by a spatiotemporal start
point ($x_i^{start}$, $y_i^{start}$, $z_i^{start}$, $t_i^{start}$), an
end point ($x_i^{end}$, $y_i^{end}$, $z_i^{end}$, $t_i^{end}$), a
segment id and a trajectory id.  Segments belonging to the same trajectory
have the same trajectory id and are ordered temporally by their segment
ids. We call $t_i^{end}-t_i^{start}$ the \emph{temporal extent} of $l_i$. 

The distance threshold search searches for 
entry segments within a distance $d$ of a query set $Q$, where $Q$ is a set of line segments that belong to a series of moving object trajectories. We call the line segments in $Q$ \emph{query segments} and denote them by $q_k, k = 1, \ldots, |Q|$. The
search is continuous, such that an entry segment may be within the distance
threshold $d$ of particular query segment for only a subinterval of that
segment's temporal extent. We call a comparison between an entry segment and a query segment an \emph{interaction}. The result set thus contains a set of query and entry segment pairs, and for each pair the time interval during which the two segments are within a distance $d$ of each other. For example, a search may return ($q_1$,$l_1$,[0.1,0.3]) and ($q_1$,$l_2$,[0.5,0.95]), for a query segment $q_1$ with temporal extent [0,1].

We consider a platform that consists of a host, with RAM and CPUs, and a
GPU with its own memory and Streaming Multi-Processors (SMPs)
connected to the CPU via a (PCI Express) bus. We consider an
\emph{in-memory database}, meaning that $D$ is stored once and for all in
global memory on the GPU, i.e., the database is stored once and queried
multiple times.  The objective is to minimize the response time for
processing the queries in $Q$. This is the typical objective considered in
other spatiotemporal database works such as the ones reviewed in
Section~\ref{sec:related_work}. We consider the case in which both $D$ and
$Q$ can fit in GPU memory. This means that GPU memory is large enough and
not shared with other users. Our intended scenario is that of a distributed
memory environment in which a number of GPU-equipped compute nodes are
reserved by a user.

\subsection{Memory Management on the GPU}
\label{sec:memmanagement}

Previous works on indexing trajectories for the purpose of distance
threshold similarity searches have targeted multi-core CPU
implementations~\cite{Gowanlock2014,Gowanlock2014b, Arumugam2006} and GPU
implementations~\cite{Gowanlock2014c}.  CPU implementations rely on
(in-memory) index trees that have been used traditionally for out-of-core
implementations, such as the R-tree~\cite{Guttman-R_tree}. Each thread
traverses the tree and creates a candidate segment set to be further
processed to create the final result set. Although many candidate segments
are not within distance $d$ of the query segments due to the ``wasted
space'' in the index (an unavoidable consequence of using
MBBs)~\cite{Gowanlock2014}, memory must still be allocated to store these
candidate segments. Furthermore, the size of the final result set is
non-deterministic as it depends on the spatiotemporal nature of the data.
Consequently, memory allocation for the result set must be conservative and overestimate
the memory required (this overestimation grows linearly with $|Q|$).
On the CPU these memory management issues are typically not problematic in practice
since the number of threads is limited (e.g., set to the number of physical
cores) and the memory is large.


On the GPU, even though we assume that both $D$ and $Q$ fit in memory, the
same memory management issues are problematic.  This is because we have a
large number of threads that each need memory to store candidate segments,
in addition to the memory needed to store the final result set.  To address
this issue of non-deterministic storage requirements, on the GPU one must
define a fixed size for a statically allocated memory buffer for each
thread. If the memory requirements exceed this buffer then it is necessary
to perform a series of kernel invocations so as to ``batch'' the generation
of the candidate sets and the final result set.

\section{Indexing Trajectory Data}\label{sec:index}

In this section we outline three trajectory indexing techniques for the
GPU.  For each we discuss shortcomings and possible solutions regarding the
memory management issues discussed in Section~\ref{sec:memmanagement}.
Although our GPU implementations use OpenCL, in what follows we use the
more common CUDA terminology to describe our algorithms (GPU as opposed to
device, kernel as opposed to program, thread as opposed to work-item,
etc.).


\subsection{Spatial Indexing: Flatly Structured Grids}

Previous work has proposed the use of grid files, or ``flatly structured
grids'' (FSG), to index trajectory data on the GPU spatially
\cite{Zhang:2012:USH:2390226.2390229}.  In that work the authors focus on
2-D spatial data (and Hausdorff distance) while our context is 3-D
spatiotemporal data (and Euclidian distance). An interesting question is
whether spatial indexing with FSGs is effective even when the data has a
temporal dimension. In what follows we describe an FSG indexing scheme and
accompanying search algorithm for the GPU. We call this approach \gpuS.

\subsubsection{Trajectory Indexing}

\begin{figure}[t]
\centering

  \begin{tikzpicture}

\def\scale{0.6}

\foreach \x in {0,1,2,3,4} {
	\foreach \y in {0,1,2,3} {
		\draw (\x*\scale,\y*\scale) rectangle (\x*\scale+1*\scale,\y*\scale+1*\scale);
	}
}

\foreach \x/\y in {1/1,2/1,1/2,2/2,1/3,2/3,4/2} {
	\draw [fill=cyan!70!black] (\x*\scale,\y*\scale) rectangle (\x*\scale+1*\scale,\y*\scale+1*\scale);
}

\foreach \a/\b/\c/\d in {2.6/1.2/1.2/3.5, 4.1/2.1/4.8/2.9} {
	\draw [fill=green!90!black,opacity=0.5] (\a*\scale,\b*\scale) rectangle (\c*\scale,\d*\scale);
	\draw [thick] (\a*\scale,\b*\scale) -- (\c*\scale,\d*\scale);
}
\node at (1.6*\scale,2.5*\scale) {{\scriptsize $l_1$}};
\node at (4.63*\scale,2.33*\scale) {{\scriptsize $l_2$}};

\end{tikzpicture}  
    \caption{2-D example rasterization of two line segment MBBs (green) to grid cells (blue) in 
    a $4\times 5$ FSG.  $l_1$: a long line segment whose MBB spans six grid cells; $l_2$: a short line segments whose MBB spans one grid cell.}
   \label{fig:mbb_grid_ex}
\end{figure}
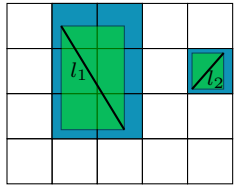

We define a FSG as a 3-D rectangular box partitioned into cells with
$grid_x$, $grid_y$, $grid_z$ cells in the $x$, $y$, and $z$ spatial
dimensions, respectively, for a total of $grid_x\times grid_y \times
grid_z$ cells. Each line segment $l_i$ in $D$ is contained in a spatial MBB
defined by two points $MBB_i^{min}$ and $MBB_i^{max}$, where $MBB_i^{min} =
(\min(x_i^{start}, x_i^{end}),\min(y_i^{start},
y_i^{end}),\min(z_i^{start}, z_i^{end}))$ and $MBB_i^{max} =
(\max(x_i^{start}, x_i^{end}),\max(y_i^{start},
y_i^{end}),\max(z_i^{start}, z_i^{end}))$.  Each line segment is assigned
to the FSG by rasterizing its MBB to grid cells.
Figure~\ref{fig:mbb_grid_ex} shows a 2-D example for two line segments and
a $5\times 5$ FSG. Each line segment may occupy more than one grid cell,
and some grid cells can remain empty.  We store the FSG as an array of
non-empty cells, $G$. Each cell is denoted as $C_h$, $h=1,\ldots,|G|$,
where $h$ is a linearized coordinate computed from the cell's $x$, $y$, and
$z$ coordinates using row-major order.


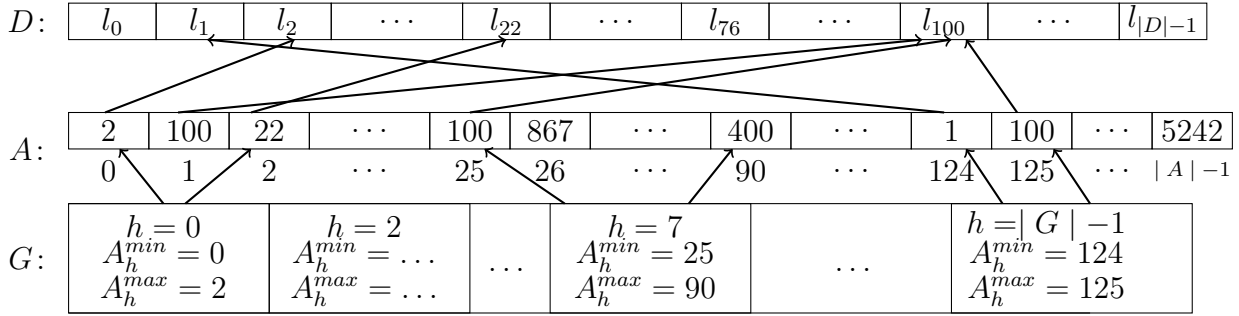
\begin{figure*}[t]
\centering
  \begin{tikzpicture}[scale=0.97]

\node at (-0.5,1.25) {$D\colon$};
\foreach \a/\b/\li in {0/1/$l_0$,1/2/$l_1$,2/3/$l_2$,3/4.5/$\ldots$,4.5/5.5/$l_{22}$,5.5/7/$\ldots$,7/8/$l_{76}$,8/9.5/$\ldots$,9.5/10.5/$l_{100}$,10.5/12/$\ldots$,12/13/$l_{\mid D\mid -1}$} {
	\draw (1.2*\a,1) rectangle (1.2*\b,1+0.5) node [pos=0.5] {\li};
}

\node at (-0.5,-0.5) {$A\colon$};
\foreach \a/\b/\li/\x in {0/1/2/0,1/2/100/1,2/3/22/2,3/4.5/$\ldots$,4.5/5.5/100/25,5.5/6.5/867/26,6.5/8/$\ldots$,8/9/400/90,9/10.5/$\ldots$,10.5/11.5/1/124,11.5/12.5/100/125,12.5/13.5/$\ldots$,13.5/14.5/5242/{\scriptsize $\mid A\mid -1$}} {
	\draw (1.1*\a,-0.5) rectangle (1.1*\b,-1.0+1) node [pos=0.5] {\li};
	\node at (1.1*\a + 1.1*0.5*\b - 1.1*0.5*\a ,-0.8) {\x};
}

\node at (-0.5,-2.25+0.25) {$G\colon$};
\foreach \a/\b/\h/\A/\B in {
                         0/2.5/$h=0$/$A_h^{min} = 0$/$A_h^{max} = 2$,
                         2.5/5/$h=2$/$A_h^{min} =\ldots$/$A_h^{max} =\ldots$,
                         6/8.5/$h=7$/$A_h^{min} = 25$/$A_h^{max} = 90$,
                         11/14/$h=\mid G\mid-1$/$A_h^{min} = 124$/$A_h^{max} = 125$} {
	\draw (1.1*\a,-3.0+0.25) rectangle (1.1*\b,-3.5+2.0+0.25);
	\node at (1.1*\a+1.3,-3.5+2.0-0.3+0.25) {\h};	
	\node at (1.1*\a+1.3,-3.5+2.0-0.7+0.25) {\A};	
	\node at (1.1*\a+1.3,-3.5+2.0-1.2+0.25) {\B};	
}
\node at (6.00,-3.5+2.0-0.9+0.25) {$\ldots$};
\node at (10.75,-3.5+2.0-0.9+0.25) {$\ldots$};
\draw (0,-3.0+0.25) -- (14,-3.0+0.25);
\draw (0,-3.5+2.0+0.25) -- (14,-3.5+2.0+0.25);

\foreach \a/\b/\c/\d in {
				1.3/-1.5+0.25/0.7/-0.5,
				1.6/-1.5+0.25/2.5/-0.5,	
				6.8/-1.5+0.25/5.7/-0.5,
				8.5/-1.5+0.25/9.1/-0.5,
				12.8/-1.5+0.25/12.3/-0.5,
				14.0/-1.5+0.25/13.5/-0.5,
				0.5/-0.0/3.1/1,
				5.5/-0.0/12.1/1,
				12/-0.0/1.9/1,
				1.5/-0.0/11.7/1,
				2.5/-0.0/6/1,
				13/-0.0/12.3/1
			} {
	\draw [->,thick] (\a,\b) -- (\c,\d);
}

\end{tikzpicture}  
    \caption{Example relationship between the grid ($G$), the lookup array ($A$) and the database of entry line segments ($D$) in the \gpuS approach.}
   \label{fig:spatial_database_layout}
\end{figure*}

Each cell $C_h$ is defined by $h$, and by an index range
$[A_h^{min}, A_h^{max}]$ in an additional integer ``lookup'' array, $A$.
$A[A_h^{min}:A_h^{max}]$ contains the indices of the line segments whose
MBBs overlap cell $C_h$ (the notation $X[a:b]$ is used to denote 
the ``slice'' of array $X$ from index $a$ to index $b$, inclusive). In other terms, if $l_i$'s MBB overlaps $C_h$,
then $i \in A[A_h^{min}:A_h^{max}]$.  Since the MBB of line segment $l_i$
can overlap multiple grid cells, $i$ can occur multiple times in array
$A$.  Figure~\ref{fig:spatial_database_layout} shows an example to
highlight the relationship between $G$, $A$, and $D$. This example is
discussed in the next section.

%


One of the objectives of the above design is to reduce the memory footprint
of the index. This is why we only index non-empty grid cells, and why for
each cell $C_h$ we do not store its spatial coordinates but instead compute
$h$ whenever needed (thereby trading off memory space for computation
time). Furthermore, the use of lookup array $A$ makes it possible for array
$G$ to consist of same-size elements (even though some cells contain more
line segments than others). Without this extra indirection through lookup
array $A$, it would have been necessary to store entry segment ids directly
into the elements of $G$. However, it would have been necessary to pick an
element size large enough to accommodate the cell with the largest number
of entry segments, thereby wasting memory space. $D$, $A$, and $G$ are
stored in GPU memory before query processing begins.

\subsubsection{Search Algorithm}

The trajectory segments in $Q$ are not sorted by any spatial or temporal
dimension. This is because sorting segments temporally would not be
effective when using a spatial index. Regarding spatial sorting, it is not
clear by which dimension the segments should be sorted.  However, segments
that are part of the same query trajectory are stored contiguously, thus providing a natural advantageous ordering of data elements.  Each
query segment $q_k$ is assigned to a GPU thread.  The kernel first
calculates the MBB for $q_k$  and the FSG cells that overlap this MBB.
Given the $x$, $y$, $z$ coordinates of each such cell in the FSG, the
kernel computes its linearized coordinate ($h$) using a row-major order. A
binary search is used to find whether cell $C_h$ occurs in array $G$, in
$O(\log |Q|)$ time.  In this manner the kernel creates a list of non-empty
cells that overlap $q_k$'s MBB. For each cell $C_h$ in this list, the
indices of the entry segments it contains are computed as
$A[A_h^{min}:A_h^{max}]$.  These indices are appended to a buffer $U_k$.  A
key point here is that with a spatial indexing scheme there is no good
approach for storing index entry segments in a contiguous manner (since one
would have to arbitrarily pick one of the spatial dimensions). This is why
we must resort to using buffer $U_k$ as opposed to, for instance, a
2-integer index range in a contiguous array of entry segments.  Each entry
in $U_k$ is then compared to the query segment $q_k$ to see if it is within
the threshold distance; however, note that while the segments are expected
to be relatively nearby each other spatially (given their FSG overlap),
they may not overlap temporally.

Consider the example in Figure~\ref{fig:spatial_database_layout}, which
shows partial contents of arrays $G$, $A$, and $D$. Consider a query $q_1$
(not shown in the figure), which overlaps grid cells $C_0$, $C_1$, and
$C_7$. Cell $C_1$ is not in $G$, meaning that it contains no entry
segments. Therefore, the only two cells to consider are $C_0$ and $C_7$,
which have [$A^h_{min}$,$A^h_{max}$] values of [0,2] and [25,90], respectively.
In lookup array $A$, we find that [0,2] corresponds to entries 2, 100, and 22,
while [25,90] corresponds to entry indices 100, 867, $\ldots$, 400.  These indices
are copied from $A$ into buffer $U_k$. Note that in this step the search
algorithm does not remove duplicate indices (such as entry index 100 in
this example) and thus may perform some redundant entry segment processing.
Removing duplicates would amount to sorting buffer $U_k$, as done for
instance in \cite{Zhang:2012:USH:2390226.2390229}, which thus comes at an
additional computational cost that may offset the benefits of removing
redundant segment processing.


Since the number of entry segments that overlap $q_k$'s MBB can be
arbitrary large (it depends on the spatial features of $D$ and $Q$, and on
the query distance $d$), the use of buffer $U_k$ creates memory pressure,
especially since both $D$ (along with $G$ and $A$) and $Q$ are stored on the GPU.  
This same issue has been encountered in previous work, e.g., when using a
parallel R-tree index on the GPU~\cite{Luo2012}.  We define an overall
buffer size, $s$, that is split equally among all queries ($|U_k| = s/|Q|$).
If the processing of query $q_k$ exceeds the capacity of $U_k$, then the
thread terminates, and stores the query id into an array that is sent
back to the host. Once the kernel execution finishes, the host re-attempts
the execution of those queries that could not complete due to memory
pressure. In this re-attempt, memory pressure is lower because fewer
queries are executed (i.e., $|U_k|$ is larger).  This method implicitly has
the effect that threads with similar (large) amounts of work to execute
together, resulting in improved load-balancing.


\newcommand{\LINEIF}[2]{%
    \State\algorithmicif\ {{#1}}\ {{#2}} %
}

\begin{algorithm}
\caption{\gpuS kernel.}
\label{alg:GPU_spatial}
\begin{algorithmic}[1]
\begin{small}
\Procedure{\gpuFSG}{$G$, $A$, $D$, $\mathbf{Q}$, {\bf queryIDs}, $U$, $d$, 
{\bf redo}, {\bf resultSet}}

\State {gid} $\leftarrow$ getGlobalId()\label{alglineFSG.pre_start}
\LINEIF{{queryIDs} $= \emptyset$ {\bf and} {gid}$\geq${$|$Q$|$}}{\Return}\label{alglineFSG.abort1}
\LINEIF{{queryIDs} $\neq \emptyset$ {\bf and} {gid}$\geq${$|$queryIDs$|$}}{\Return}\label{alglineFSG.abort2}


\If {{queryIDs}  $= \emptyset$}
\State {queryID} $\leftarrow$ {gid} \label{alglineFSG.getquery}
\Else
\State {queryID} $\leftarrow$ {queryIDs[gid]} \label{alglineFSG.getqueryReattempt}
\EndIf
\State ({overflow} , {candidateSet}) $\leftarrow$ getCandidates($G$, $A$, $D$, $Q$[{queryID}], {$U$}, {d}) \label{alglineFSG.calcoverlap_entries}
\If {overflow} \label{alglineFSG.overflow}
\State {\bf atomic:} redo $\leftarrow$ redo $\cup$ \{ queryID \} 
\State \Return\label{alglineFSG.overflow_return}
\EndIf
\ForAll {entryID $\in$ candidateSet}\label{alglineFSG.loop}
\State result $\leftarrow$ compare($D$[entryID],$Q$[{queryID}])\label{alglineFSG.compare}
\If {result $\neq$ $\emptyset$}
\State {\bf atomic:} {resultSet} $\leftarrow$ {resultSet} $\cup$ result\label{alglineFSG.found}
\EndIf
\EndFor
\State \Return\label{alglineFSG.return}
\EndProcedure
\end{small}
\end{algorithmic}
\end{algorithm}

The pseudo-code of the search algorithm is shown in
Algorithm~\ref{alg:GPU_spatial}.  It takes the following
arguments: (i)~the FSG array ($G$); (ii)~the lookup array ($A$); (iii)~the
database ($D$); (iv)~the set of queries ($Q$); (v)~an array that contains
the ids of the queries to be reprocessed (\emph{queryIDs}), which is empty
for the first kernel invocation; (vi)~buffer space
($U$); (vii)~the query distance ($d$); (viii)~an output array in which the
kernel stores the ids of the queries that must be reprocessed
(\emph{redo}); and (ix)~the memory space to
store the result set (\emph{resultSet}). Arguments that lead to array
transfers betwen the host and the GPU, either as input or output,
are shown in boldface.  Other arguments are either pointers to
pre-allocated zones of (global) GPU memory or integers.  The algorithm
begins by checking the global thread id and aborts if it is greater than
$Q$ or $|$queryIDs$|$, depending on whether this is a first invocation or
a re-invocation (lines~\ref{alglineFSG.abort1}-\ref{alglineFSG.abort2}).
The id of the query assigned to the GPU thread is then acquired from $Q$ or
using an indirection via \emph{queryIDs}
(lines~\ref{alglineFSG.getquery}-\ref{alglineFSG.getqueryReattempt}).
Function \emph{getCandidates} searches the FSG and returns a boolean that
indicates whether buffer space was exceeded and the (possibly empty) set of
candidate entry segment ids (line~\ref{alglineFSG.calcoverlap_entries}).  If
buffer space was exceeded, then the query id is atomically added to the
\emph{redo} array and the thread terminates
(line~\ref{alglineFSG.overflow}-\ref{alglineFSG.overflow_return}).  The
algorithm then loops over all candidate entry segment ids
(line~\ref{alglineFSG.loop}), compares each entry segment spatially and
temporally to the query (line~\ref{alglineFSG.compare}) and atomically
adds a query result, if any, to the result set
(line~\ref{alglineFSG.found}).  Once all GPU threads have completed,
\emph{resultSet} and \emph{redo} are transferred back to the host.  If
$|$redo$|$ is non-zero, then the kernel is re-invoked, passing \emph{redo}
as \emph{queryIDs}.  Duplicates in the result set are filtered out on the
host.

\subsection{Temporal Indexing}\label{sec:temporal_indexing}

In this section we propose a purely temporal partitioning strategy, which
we call \gpuT.

\subsubsection{Trajectory Indexing}

We begin by sorting the entries in $D$ by ascending $t_{start}$ values,
re-numbering the entry segments in this order, i.e., $t_i^{start} \leq
t_{i+1}^{start}$.   The full temporal extent of $D$ is [$t_{min},t_{max}$]
where $t_{min} =\min_{l_i \in D} t_i^{start}$ and $t_{max} = \max_{l_i \in
D} t_i^{end}$. We divide this full temporal extent so as to create $m$
logical bins of fixed length $b =(t_{max} - t_{min}) / m$.  We assign each
entry segment, $l_i$, $i=1,\ldots,|D|$, to a bin, where $l_i$ belongs to
bin $B_j$, $j=1,\ldots,m$, if $\lfloor t_i^{start} / b \rfloor = j$.  There
can be temporal overlap between the line segments in adjacent bins.  For
each bin $B_j$ we defined its start times as $B_{j}^{start} = j\times b$
and its end time as $B_j^{end} = \max((j+1)\times b, \max_{l_i \in B_j}
t_i^{end})$. $B_{j}^{start}$ does not depend on the line segments in bin
$B_j$, but $B_{j}^{end}$ does.  The temporal extent of bin $B_j$ is
defined as [$B_{j}^{start},B_{j}^{end}$]. Given the definitions of
$B_{j}^{start}$ and $B_{j}^{end}$, the union of the temporal extents of the
bins is equal to the full temporal extent of $D$.  We define $B_j^{first} =
\arg\min_{i | l_i \in B_j} t_i^{start}$ and $B_j^{last} = \arg\max_{i | l_i
\in B_j} t_i^{start}$, i.e., the ids of the first and last entry segments
in bin $B_j$, respectively.  [$B_{j}^{first},B_{j}^{last}$] forms the index
range of the entry segments in $B_j$.  Bin $B_j$ is thus fully described as
($B_{j}^{start}$,$B_{j}^{end}$, $B_{j}^{first},B_{j}^{last}$).  The set of
bins forms the temporal database index.

\begin{figure}[t]
\centering
  \begin{tikzpicture}

  \def\xscale{1.2}

  \draw [->] (0,0) -- (\xscale*13.5,0);
  \node (timelabel) at (\xscale*13.0,-0.25) {\small{time}};
  \foreach \x in {0,...,4} {
  	\draw [dashed] (\xscale*3*\x,-0.1) -- (\xscale*3*\x,3.48);
  }
  \foreach \x in {0,...,12} {
	\draw (\xscale*\x,0) -- (\xscale*\x,-0.1);
	\node [label=south:\footnotesize{\x}] at (\xscale*\x , -0.0) {};
  }

\def\bins{
	0/0/7.5/0/5,
	1/3/6.2/6/8,
	2/6/11/9/11,
	3/9/12/12/14
	}
  \foreach \x/\a/\b/\c/\d in \bins {
	\def\height{3.8}
	\def\increment{0.45}
	\node [label=south:\scriptsize{bin $B_{\x}$}] at (\xscale*1.5+\xscale*3*\x , \height) {};
	\node [label=south:\scriptsize{($B_{\x}^{start},B_{\x}^{end})=(\a,\b)$}] at (\xscale*1.5+\xscale*3*\x , \height-\increment) {};
	\node [label=south:\scriptsize{($B_{\x}^{first},B_{\x}^{last})=(\c,\d)$}] at (\xscale*1.5+\xscale*3*\x , \height-2*\increment) {};
  }

  \def\positions{
	0.0/1.7/0.2/0/north,
 	0.6/1.5/0.4/1/south,
	0.7/1.3/0.6/2/north,
	1.9/3.7/0.4/3/north,
 	2.5/7.5/0.6/4/north,
	2.8/4.6/0.2/5/south,
	3.9/5.5/0.4/6/south,
 	4.1/5.8/0.8/7/north,
	4.8/6.2/0.2/8/south,
	6.5/9.4/0.4/9/south,
	7.0/7.8/0.2/10/south,
	8.3/11/0.2/11/south,
	9.3/11.5/0.6/12/north,
	10.4/12/0.4/13/south,
	11.6/11.9/0.2/14/south}

  \def\yscale{2.10}
  \def\yoffset{0.10}
  \tikzstyle{every node} = [draw,fill=black,inner sep=1.0pt]
  \foreach \a/\b/\y/\z/\p in \positions {
	\ifthenelse{\equal{\p}{south}}{
  	\node [label=south:\footnotesize{$l_{\z}$}] (start) at (\xscale*\a,\y*\yscale+\yoffset) {}; 
        }{
  	\node [label=north:\footnotesize{$l_{\z}$}] (start) at (\xscale*\a,\y*\yscale+\yoffset) {}; 
        }
        \node (end) at (\xscale*\b,\y*\yscale+\yoffset) {}; 
        \draw (start) -- (end);
  }

\end{tikzpicture}  
    \caption{An example assignment of entry line segments to temporal bins in the \gpuT approach.}
   \label{fig:example_index}
\end{figure}
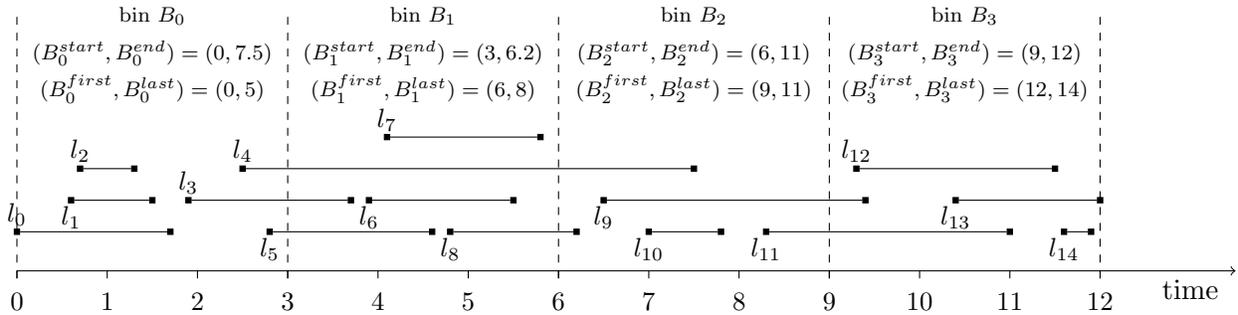

Figure~\ref{fig:example_index} shows an example of how line segments may be
assigned to a set of temporal bins. In this example, 15 entry segments
are assigned to 4 temporal bins over a database temporal extent of 12
time units (spatial dimensions are ignored, and thus line segments are
simply represented as horizontal lines in the figures).  The
$B_{}^{start}$,$B_{}^{end}$, $B_{}^{first}$ and $B_{}^{last}$ values are
shown for each bin.  For instance, three entry segments are assigned to bin
$B_2$: $l_9$, $l_{10}$, and $l_{11}$.  Thus, $B_{2}^{first} = 9 $ and
$B_{2}^{last} = 11$. $B_{2}^{start}=2\times(12/4) = 6$ and $B_{2}^{end} =
t_{11}^{end}$.

\subsubsection{Search Algorithm}

Before performing the actual search, the following pre-processing steps
must be performed.  First, query segments in $Q$ are sorted by
non-decreasing $t_{start}$ values, in $O(|Q| \log |Q|)$ time. For each
query segment $q_k$, we calculate the index range of the contiguous bins
that it overlaps temporally. A na\"ive algorithm for computing this overlap
would be to scan all bins in $O(m)$ time. A binary search could be used to
obtain a logarithmic time complexity.  In practice, however, there are many
temporally contiguous query segments and each overlaps only a few bins.
Since segments in $Q$ are sorted by non-decreasing $t_{start}$, the search
can be done efficiently by using the first temporal bin that overlaps the
previous query segment as the starting point for the scan for the temporal
bins that overlap the next query segment.  The search thus typically takes
near-constant time. Let $\mathcal{B}_k$ denote the set of contiguous bins
that temporally overlap query segment $q_k$, as identified by the above
search. In constant time we can now compute the index range of the entry
line segments that may overlap and must be compared with $q_k$: $E_k =
[\min_{B\in \mathcal{B}} B_{j}^{first}, \max_{B\in \mathcal{B}}
B_{j}^{last}]$.  We term the mapping between $q_k$ and $E_k$ the
\emph{schedule}, $S$. Each GPU thread compares a single query to the line
segments in $D$ whose indices are in the $E_k$ range.  Assuming that $|Q|$
is moderately large, one is then insured that all GPU cores can be
utilized.

In our implementation, all preprocessing described in the previous
paragraph is performed on the CPU. Some of this preprocessing could be
performed on the GPU (e.g., sorting the query segments). In an initial
implementation, we performed the calculation of $E_k$ on the GPU; however,
this did not result in any performance improvement.  As explained earlier,
on the host the search for temporally overlapping bins can be drastically
improved by relying on the same search for the previous query segment.
However, this cannot be implemented on the GPU as it would require thread
synchronization and communication, which cannot be performed across thread
blocks.  In all of our experiments, the time to compute $S$ on the CPU is a
negligible portion of the overall query response time.

\begin{algorithm}
\caption{\gpuT kernel.}
\label{alg:GPU_temporal}
\begin{algorithmic}[1]

\begin{small}
\Procedure{\gputemporal}{$D$, {$\mathbf{Q}$}, {$\mathbf{S}$}, $d$, {\bf resultSet}}

\State {gid} $\leftarrow$ getGlobalId()\label{algline.pre_start}
\LINEIF{{gid}$\geq${$|Q|$}}{\Return}\label{algline.abort}

\State {queryID} $\leftarrow$ gid \label{algline.queryseg_into_private_memory}

\State {entryMin} $\leftarrow$ $S$[gid].EntryMin\label{algline.entryidmin_into_private_memory}
\State {entryMax} $\leftarrow$ $S$[gid].EntryMax\label{algline.entryidmax_into_private_memory}

\ForAll {entryID $\in$ $\{$entryMin,...,entryMax$\}$}\label{algline.loop}


%
%
\State result $\leftarrow$ compare($D$[entryID],$Q$[queryID])\label{algline.compare}
\If {result $\neq$ $\emptyset$}
\State {\bf atomic:} {resultSet} $\leftarrow$ {resultSet} $\cup$ result\label{algline.found}
\EndIf
\EndFor\label{algline.endloop}
\State \Return\label{algline.return}
\EndProcedure

\end{small}
\end{algorithmic}
\end{algorithm}

The pseudo-code of the search algorithm is shown in
Algorithm~\ref{alg:GPU_temporal}.  It takes the following arguments:
(i)~the database ($D$); (ii)~the query set ($Q$); (iii)~the schedule ($S$);
(iv)~the query distance ($d$); and (v)~the memory space to store the result
set (\emph{resultSet}).  As in Algorithm~\ref{alg:GPU_spatial}, arguments
that lead to array transfers between the host and the GPU are shown in
boldface.  The algorithm first checks the global thread id and aborts if it
is greater than $|Q|$ (line~\ref{algline.abort}).  The query assigned to
the thread is then acquired from $Q$
(line~\ref{algline.queryseg_into_private_memory}).  Next, the algorithm
retrieves the minimum and maximum entry segment indices from the schedule
(lines~\ref{algline.entryidmin_into_private_memory}-\ref{algline.entryidmax_into_private_memory}).
From line~\ref{algline.loop} to \ref{algline.return} the algorithm then operates as Algorithm~\ref{alg:GPU_spatial}.

\subsection{Spatiotemporal Indexing}
\label{sec:spatiotemporal_indexing}

In the two previous sections we have proposed a purely spatial and a purely
temporal indexing scheme.  The spatial scheme leads to segments in $Q$ and
$D$ being compared that are spatially relevant but may be temporal misses
(no temporal overlap).  Likewise, the temporal indexing scheme compares
temporally relevant segments in $Q$ and $D$, but these segments may be
spatial misses (no spatial overlap). Therefore, either approach can
outperform the other depending on the spatiotemporal characteristics of $Q$
and $D$. Assuming for the sake of discussion that these characteristics do
not give any such particular advantage to either one of the two indexing
approaches, we can reason about their relative performance.  First, the spatial
indexing approach requires buffer space to store the spatially overlapping
trajectory segments. In contrast, because the temporal indexing scheme is
indexed in a single dimension, the temporally overlapping entry segments
can be defined by an index range in $D$, which represents significant
memory space savings. The same method could possibly be used with a spatial
indexing scheme if considering only one of the spatial dimensions, making
the index no longer a multi-dimensional grid, but instead a linear array.
This approach would however drastically decrease the spatial selectivity of
the search, leading to large increases in wasted computational effort
(i.e., comparisons of segments that have no overlap in one or two of the
spatial dimensions).  Second, to minimize the memory footprint on the GPU,
the spatial scheme requires two additional arrays ($G$ and $A$), thus
leading to two indirections in global GPU memory. In contrast, the temporal
scheme requires a single indirection.  Moreover, the entry segments are
stored contiguously in the temporal scheme, while this is not the case in
the spatial scheme.

Given the features of both the spatial and the temporal indexing scheme, we
attempt to find an alternate spatiotemporal index that retains the benefit
of both schemes without some of the drawbacks mentioned above. We term
this approach \gpuST.

\subsubsection{Trajectory Indexing}

\gpuST adopts a temporal index so as to avoid the buffering and multiple
indirection issues of spatial indexing, but subdivides each temporal bin
into spatial subbins to achieve spatial selectivity.  Entry segments in $D$
are assigned to $m$ temporal bins exactly as for \gpuT.  We then compute
the spatial extent of $D$ in each dimension. For instance, in the $x$
dimension the extent of $D$ is: $$[x_{min},x_{max}] = [\min_{l_i \in
D}(\min(x^i_{start},x^i_{end})), \max_{l_i \in
D}(\max(x^i_{start},x^i_{end}))]\;.$$ Spatial extents in the $y$ and $z$
dimensions are computed similarly.  We then compute the maximum spatial
extent in each dimension of the entry segments, which for the $x$
dimensions is $\max_{l_i \in D} |x^i_{start} - x^i_{end}|$.  Maximum
spatial extents are computed similarly for the $y$ and $z$ dimension.  For
each of the temporal bins, we create $v$ spatial subbins along each
dimension, with the constraints that these subbins are larger than the
maximum spatial extent of the entry segments.  For instance, in the $x$
dimension, this constraint is expressed as $v \leq (x_{max} - x_{min}) /
\max_{l_i \in D} |x^i_{start} - x^i_{end}|$.  We place this constraint for
two reasons, which will be clarified when we describe the search algorithm:
(i)~to eliminate duplicates in the result set, and (ii)~to reduce the
amount of redundant information in the index.  In total we have $m\times v$
subbins and we denote each subbin as $\hat{B}_{i,j}$, with $i=1,\ldots,m$
and $j=1,\ldots,v$.

The part of Figure~\ref{fig:spatiotemporal_indexing} above the dashed line 
shows an example of how entry line segments are logically assigned to bins and subbins.
The very top of the figure shows $m=3$ temporal bins, $B_0$ to $B_2$. Each
temporal bin contains contains the segments with ids in the range
$[B_j^{first},B_j^{last}]$.  For instance, $B_1^{first} = 4$ and
$B_1^{last} = 7$.  Each entry segment is described by an id and 2 spatial
$(x,y,z)$ extremities.  For instance, segment $l_6$ is
in temporal bin $B_1$ and its spatial extremities are $(8,9,10)$ and
$(10,9,8)$.  Temporal dimensions are omitted in the figure.  Below the
temporal bins, we depict 9 temporal spatial subbins, $\hat{B}_{0,0}$ to
$\hat{B}_{2,2}$ ($v=3$ subbins per temporal bin). For each subbin, we indicate
its spatial range in the $x$, $y$, and $z$ dimension. Each subbin spans 4
spatial units in the $x$ and $y$ dimensions, and 5 spatial units in the $z$
dimension. Given segment lengths in the database these subbin dimensions
meet the constraints described in the previous paragraph. For each subbin
and each dimension, we show the overlapping entry segment ids. For instance,
consider subbin $\hat{B}_{0,1}$. It is overlapped in the $x$ dimension by $l_0$,
$l_2$ and $l_3$, in the $y$ dimension by $l_3$, and in the $z$
dimension by $l_1$ and $l_3$. 

The part of Figure~\ref{fig:spatiotemporal_indexing} below the dashed line shows how
the logical assignment of segments to spatial subbins is implemented 
physically in memory.  We create three integer arrays, $X$, $Y$, and $Z$, 
depicted at the bottom of the figure.
Each array stores the ids of the line segments that overlap the subbins in
one spatial dimension. The ids for a subbin are stored contiguously,
for the subbins $\hat{B}_{i,j}$'s sorted by $(j,i)$ lexicographical order. 
This is illustrated using colors in the figure
and amounts to storing contiguously all ids in the first subbins of the
temporal bins, then all ids in the second subbins of the temporal bins,
etc. For instance, for the $y$ dimension, the $Y$ array in our example
consists of $v=3$ chunks. The first chunk corresponds to the ids in subbins
$\hat{B}_{0,0}$ ($l_0$,$l_1$,$l_3$), $\hat{B}_{1,0}$ ($l_4$, $l_5$, $l_8$), and
$\hat{B}_{2,0}$ ($l_9$), the second chunk corresponds to the ids in subbins
$\hat{B}_{0,1}$ ($l_3$), $\hat{B}_{1,1}$ ($l_5$, $l_7$), and $\hat{B}_{2,1}$ ($l_8$),
and the third chunk corresponds to the ids in subbins $\hat{B}_{0,2}$ (none),
$\hat{B}_{1,2}$ ($l_6$), and $\hat{B}_{2,2}$ ($l_8$).  
The reason for storing the ids in this manner is as follows. 
Consider a query segment with some spatial and temporal extent. This query
may overlap several contiguous temporal bins (as shown in
Section~\ref{sec:temporal_indexing}). However, because of the way in which
we choose the sizes of the spatial subbins, most queries will not overlap
multiple subbins in all three dimensions. Identifying potential overlapping
entry segments then amounts to examining the $i$-th subbin of contiguous
temporal bins, for some $0 \leq i \leq v$. In other words, based on the example
in Figure~\ref{fig:spatiotemporal_indexing}, this amounts to examining sequences of same-color
subbins. 

Given the $X$, $Y$, and $Z$ array, each spatial subbin is then described
with the index range of the entries in those arrays, i.e., 6 integers.  For
instance, consider subbin $\hat{B}_{0,1}$ in our example. Its description is
index range $5-7$ in the $x$ dimension (i.e., it overlaps with segments
$l_{X[5]}$ to $l_{X[7]}$ in the $x$ dimension), index range $7-7$ in the
$y$ dimension (i.e., it overlaps with segment $l_{X[7]}$ in the $y$
dimension), and index range $4-5$ in the $z$ dimension (i.e., it overlaps
with segment $l_{X[4]}$ $l_{X[5]}$ in the $z$ dimension). Using 
this indirection, each spatial subbin is of fixed size.  When compared
to the purely temporal index, this spatiotemporal indexing scheme 
requires only additional space in GPU memory for the $X$, $Y$, and $Z$
integer arrays, which corresponds to $\gtrsim 3|D| \times 4$ bytes.

\begin{figure}[t]
\centering
  \begin{tikzpicture}

\definecolor{ao}{rgb}{0.0, 0.5, 0.0}

\def\firstyoffset{0.9}
\def\xoffset{-0.4}
\def\square{0.35}
\node at (\xoffset+0+0.2,0+\firstyoffset) {$X$:};
\foreach \x/\v/\c in { 0/1/red, 1/3/red, 2/4/red, 3/5/red, 4/9/red, 5/0/cyan, 6/2/cyan, 7/3/cyan, 8/4/cyan, 9/7/cyan, 10/8/cyan, 11/6/ao, 12/8/ao}  {
  \draw (\xoffset+0.5+\square*\x,-0.5*\square+\firstyoffset) rectangle(\xoffset+0.5+\square*\x+\square,0.5*\square+\firstyoffset);
  \node at (\xoffset+0.5+\square*\x+0.5*\square,0+\firstyoffset) {{\scriptsize {\bf \textcolor{\c}{\v}}}};
  \node at (\xoffset+0.5+\square*\x+0.5*\square,-0.35+\firstyoffset) {{\tiny {\bf \x}}};
}

\def\xoffset{5.1}
\def\square{0.35}
\node at (\xoffset+0+0.2,0+\firstyoffset) {$Y$:};
\foreach \x/\v/\c in { 0/0/red, 1/1/red, 2/3/red, 3/4/red, 4/5/red, 5/8/red, 6/9/red, 7/3/cyan, 8/5/cyan, 9/7/cyan, 10/8/cyan, 11/6/ao, 12/8/ao}  {
  \draw (\xoffset+0.5+\square*\x,-0.5*\square+\firstyoffset) rectangle(\xoffset+0.5+\square*\x+\square,0.5*\square+\firstyoffset);
  \node at (\xoffset+0.5+\square*\x+0.5*\square,0+\firstyoffset) {{\scriptsize {\bf \textcolor{\c}{\v}}}};
  \node at (\xoffset+0.5+\square*\x+0.5*\square,-0.35+\firstyoffset) {{\tiny {\bf \x}}};
}

\def\xoffset{10.6}
\def\square{0.35}
\node at (\xoffset+0+0.2,0+\firstyoffset) {$Z$:};
\foreach \x/\v/\c in { 0/0/red, 1/1/red, 2/4/red, 3/5/red, 4/1/cyan, 5/3/cyan, 6/4/cyan, 7/6/cyan, 8/7/cyan, 9/8/cyan, 10/6/ao, 11/8/ao}  {
  \draw (\xoffset+0.5+\square*\x,-0.5*\square+\firstyoffset) rectangle(\xoffset+0.5+\square*\x+\square,0.5*\square+\firstyoffset);
  \node at (\xoffset+0.5+\square*\x+0.5*\square,0+\firstyoffset) {{\scriptsize {\bf \textcolor{\c}{\v}}}};
  \node at (\xoffset+0.5+\square*\x+0.5*\square,-0.35+\firstyoffset) {{\tiny {\bf \x}}};
}

\def\yoffset{0.6}
\def\xsquare{1.6}
\def\ysquare{1.4}

\node at (0,\yoffset+1.1+0.4) {Lookup};
\node at (0,\yoffset+0.7+0.4) {~Ids:};

\foreach \x/\c/\A/\B/\C in {
0/red/$X$:0-1/$Y$:0-2/$Z$:0-1,
1/cyan/$X$:5-7/$Y$:7-7/$Z$:4-5,
2/ao/$X$:$\varnothing$/$Y$:$\varnothing$/$Z$:$\varnothing$,
3/red/$X$:2-3/$Y$:3-5/$Z$:2-3,
4/cyan/$X$:8-9/$Y$:8-9/$Z$:6-8,
5/ao/$X$:11-11/$Y$:11-11/$Z$:10-10,
6/red/$X$:4-4/$Y$:6-6/$Z$:$\varnothing$,
7/cyan/$X$:10-10/$Y$:10-10/$Z$:9-9,
8/ao/$X$:12-12/$Y$:12-12/$Z$:11-11} {

\draw [fill=\c] (1.0+\xsquare*\x,\yoffset+0.2+0.2+0.4) rectangle (1.0+\xsquare*\x+\xsquare,\yoffset+\ysquare+0.4);
\node [anchor=west] at (1.0+\xsquare*\x+0.0,\yoffset+\ysquare-0.2+0.4) {{\small \A}};
\node [anchor=west] at (1.0+\xsquare*\x+0.0,\yoffset+\ysquare-0.5+0.4) {{\small \B}};
\node [anchor=west] at (1.0+\xsquare*\x+0.0,\yoffset+\ysquare-0.8+0.4) {{\small \C}};
}

\draw [dashed,thick] (-0.5,\yoffset+1.7+0.4) -- (15.5,\yoffset+1.7+0.4);

\def\yyoffset{2}
\node at (0,\yoffset+\yyoffset+1.0+0.1) {Entry};
\node at (0,\yoffset+\yyoffset+0.6+0.1) {~Ids:};

\foreach \x/\c/\A/\B/\C in {
0/red/$x$: 1{,}3/$y$: 0{,}1{,}3/$z$: 0{,}1,
1/cyan/$x$: 0{,}2{,}3/$y$: 3/$z$: 1{,}3,
2/ao/$x$:/$y$:/$z$:,
3/red/$x$: 4{,}5/$y$: 4{,}5{,}8/$z$: 4{,}5,
4/cyan/$x$: 4{,}7/$y$: 5{,}7/$z$: 4{,}6{,}7,
5/ao/$x$: 6/$y$: 6/$z$: 6,
6/red/$x$: 9/$y$: 9/$z$:,
7/cyan/$x$: 8/$y$: 8/$z$: 9,
8/ao/$x$: 8/$y$: 8/$z$: 8} {

\draw [] (1.0+\xsquare*\x,\yoffset+\yyoffset+0.4) rectangle (1.0+\xsquare*\x+\xsquare,\yoffset+\yyoffset+\ysquare);
\node [anchor=west] at (1.0+\xsquare*\x+0.0,\yoffset+\yyoffset+\ysquare-0.2) {\textcolor{\c}{{\small \A}}};
\node [anchor=west] at (1.0+\xsquare*\x+0.0,\yoffset+\yyoffset+\ysquare-0.5) {\textcolor{\c}{{\small \B}}};
\node [anchor=west] at (1.0+\xsquare*\x+0.0,\yoffset+\yyoffset+\ysquare-0.8) {\textcolor{\c}{{\small \C}}};
}

\def\yyyoffset{1.5}
\node at (0,\yoffset+\yyoffset+\yyyoffset+0.9) {Spatial};
\node at (0,\yoffset+\yyoffset+\yyyoffset+0.4) {~ranges:};
\node at (0,\yoffset+\yyoffset+\yyyoffset+1.8) {Subbins:};

\foreach \x/\c/\A/\B/\C/\bin in {
0/black/$x$:[0{,}4)/$y$:[0{,}4)/$z$:[0{,}5)/$\hat{B}_{0,0}$,
1/black/$x$:[4{,}8)/$y$:[4{,}8)/$z$:[5{,}10)/$\hat{B}_{0,1}$,
2/black/$x$:[8{,}12)/$y$:[8{,}12)/$z$:[10{,}15)/$\hat{B}_{0,2}$,
3/black/$x$:[0{,}4)/$y$:[0{,}4)/$z$:[0{,}5)/$\hat{B}_{1,0}$,
4/black/$x$:[4{,}8)/$y$:[4{,}8)/$z$:[5{,}10)/$\hat{B}_{1,1}$,
5/black/$x$:[8{,}12)/$y$:[8{,}12)/$z$:[10{,}15)/$\hat{B}_{1,2}$,
6/black/$x$:[0{,}4)/$y$:[0{,}4)/$z$:[0{,}5)/$\hat{B}_{2,0}$,
7/black/$x$:[4{,}8)/$y$:[4{,}8)/$z$:[5{,}10)/$\hat{B}_{2,1}$,
8/black/$x$:[8{,}12)/$y$:[8{,}12)/$z$:[10{,}15)/$\hat{B}_{2,2}$}  {

\draw [] (1.0+\xsquare*\x,\yoffset+\yyoffset+\yyyoffset) rectangle (1.0+\xsquare*\x+\xsquare,\yoffset+\yyoffset+\yyyoffset+\ysquare);
\node [anchor=west] at (1.0+\xsquare*\x+0.0,\yoffset+\yyoffset+\yyyoffset+\ysquare-0.3) {\textcolor{\c}{{\small \A}}};
\node [anchor=west] at (1.0+\xsquare*\x+0.0,\yoffset+\yyoffset+\yyyoffset+\ysquare-0.7) {\textcolor{\c}{{\small \B}}};
\node [anchor=west] at (1.0+\xsquare*\x+0.0,\yoffset+\yyoffset+\yyyoffset+\ysquare-1.1) {\textcolor{\c}{{\small \C}}};

\node at (1.5+\xsquare*\x+0.5*\square,\yoffset+\yyoffset+\yyyoffset+1.8) {\bin};
}

\def\yyyyoffset{-0.2}

\draw (1.0,\yoffset+\yyoffset+\yyyoffset+\yyyyoffset+2.4) rectangle (1.0+3*\xsquare,\yoffset+\yyoffset+\yyyoffset+\yyyyoffset+2.5+2-0.4);
\draw (1.0+3*\xsquare,\yoffset+\yyoffset+\yyyyoffset+\yyyoffset+2.4) rectangle (1.0+6*\xsquare,\yoffset+\yyoffset+\yyyoffset+\yyyyoffset+2.5+2-0.4);
\draw (1.0+6*\xsquare,\yoffset+\yyoffset+\yyyoffset+\yyyyoffset+2.4) rectangle (1.0+9*\xsquare,\yoffset+\yyoffset+\yyyoffset+\yyyyoffset+2.5+2-0.4);

\node [anchor=west] at (1,\yoffset+\yyoffset+\yyyoffset+\yyyyoffset+2.5+2-0.3-0.4) {$l_0$: (4,2,3) (5,3,1)};
\node [anchor=west] at (1,\yoffset+\yyoffset+\yyyoffset+\yyyyoffset+2.5+2-0.8-0.3) {$l_1$: (2,3,4) (1,2,2)};
\node [anchor=west] at (1,\yoffset+\yyoffset+\yyyoffset+\yyyyoffset+2.5+2-1.3-0.2) {$l_2$: (6,7,9) (4,6,8)};
\node [anchor=west] at (1,\yoffset+\yyoffset+\yyyoffset+\yyyyoffset+2.5+2-1.8-0.1) {$l_3$: (3,5,4) (4,3,5)};

\node [anchor=west] at (1+3*\xsquare,\yoffset+\yyoffset+\yyyyoffset+\yyyoffset+2.5+2-0.3-0.4) {$l_4$: (3,3,1) (5,3,7)};
\node [anchor=west] at (1+3*\xsquare,\yoffset+\yyoffset+\yyyyoffset+\yyyoffset+2.5+2-0.8-0.3) {$l_5$: (3,6,2) (2,3,2)};
\node [anchor=west] at (1+3*\xsquare,\yoffset+\yyoffset+\yyyyoffset+\yyyoffset+2.5+2-1.3-0.2) {$l_6$: (8,9,10) (10,9,8)};
\node [anchor=west] at (1+3*\xsquare,\yoffset+\yyoffset+\yyyyoffset+\yyyoffset+2.5+2-1.8-0.1) {$l_7$: (5,5,6) (6,4,5)};

\node [anchor=west] at (1+6*\xsquare,\yoffset+\yyoffset+\yyyoffset+\yyyyoffset+2.5+2-0.3-0.4) {$l_8$: (8,8,13) (7,7,10)};
\node [anchor=west] at (1+6*\xsquare,\yoffset+\yyoffset+\yyyoffset+\yyyyoffset+2.5+2-0.8-0.3) {$l_9$: (0,3,5) (2,6,7)};

\node at (0,\yoffset+\yyoffset+\yyyoffset+\yyyyoffset+3.5-0.2) {Entries:};
\node at (0,\yoffset+\yyoffset+\yyyoffset+\yyyyoffset+4.8-0.4) {Bins:};
\node at (1+1.5*\xsquare,\yoffset+\yyoffset+\yyyoffset+\yyyyoffset+4.8-0.4) {$B_0$};
\node at (1+3*\xsquare+1.5*\xsquare,\yoffset+\yyoffset+\yyyoffset+\yyyyoffset+4.8-0.4) {$B_1$};
\node at (1+6*\xsquare+1.5*\xsquare,\yoffset+\yyoffset+\yyyoffset+\yyyyoffset+4.8-0.4) {$B_2$};

\end{tikzpicture}
    \caption{Example spatiotemporal indexing of a dataset with 10 entry
    segments. Above the dashed line is the logical assignment of the
    segments to the spatial subbin. Below the dashed line is the physical
    realization of this assignment in GPU memory.}
   \label{fig:spatiotemporal_indexing}
\end{figure}

\subsubsection{Search Algorithm}

On the host, as in the \gpuT approach, we first
sort $Q$ and for each query segment calculate the
temporally overlapping entries from the temporal bins.
We also compute the
set of spatially overlapping subbins in each dimension. This computation
also takes place on the host, where the description of the bins and subbins are
stored. Arrays $X$, $Y$, and $Z$ are stored on the GPU.  One option would
be to compute the intersection of entry segments that belong to these
subbins so as to select only spatially relevant entry segments. This turns
out to be inefficient because we would then have to send a list of entry
segment indices to the GPU, which has high overhead. Instead, we seek a
solution in which we send a fixed and small number of indices to the GPU.
As a result, we opt for a poorer but easier to encode selection of the
candidate entry segments.  Among the three spatial dimensions we pick the
one in which the number of entry segments that overlap the query segment is
the smallest. We then simply send an index range, 2 integers, in the $X$,
$Y$, or $Z$ array, depending on the dimension that was picked.  This
approach may lead to wasteful computation on the GPU (i.e., evaluation of
entry segments that do not overlap with the query segment in one of the
other two spatial dimensions), but the overhead of these wasteful
computations is offset by the gain from the reduced amount of data that is
sent to the GPU. Let us demonstrate how this approach exploits the way in
which the $X$, $Y$, and $Z$ arrays are constructed in the previous section.
For the example in Figure~\ref{fig:spatiotemporal_indexing}, consider a
query segment that overlaps temporal bins 0 and 1, and overlaps spatially
with subbins $\hat{B}_{0,0}$ and $\hat{B}_{1,0}$ in the $x$ dimension
(entries 1,3,4,5), with subbins $\hat{B}_{0,1}$ and $\hat{B}_{1,2}$ in the
$y$ dimension (entries 3,5,7), and with subbins $\hat{B}_{0,0}$ and
$\hat{B}_{1,0}$ in the $z$ dimension (entries 0,1,4,5).  Because the smallest number
of entries in the overlapping subbins is along the $y$ dimension, we opt to
compare the query with entries 3,5, and 7.  In array $Y$, these entries
are stored \emph{contiguously} at indices 7, 8, and 9. So we simply compare
the query to the entry segments stored in array $Y$ from index 7 to index
9, which is encoded as one dimension specification and two integers, i.e.,
a constant size w.r.t. to the number of entry segments.  We perform a
comparison of the query segment with entry segment 7, even though entry 7
does not overlap the query along the $x$ and $z$ dimension. This comparison
will thus lead to wasteful computation due to our non-perfect spatial
selectivity of entry segments.


On the host, we generate a schedule $S$, which contains for each query
segment $q_k$ a specification of which lookup array to use (0 for $X$, 1
for $Y$, or 2 for $Z$) and an index range into that array, which we encode
using 4 integers (which preserves alignment).  \gpuST requires only 1 extra
indirection in comparison to \gpuT, and avoids storing the overlapping
entry indices in a buffer like in \gpuS.  We then sort $S$ based on the
lookup array specification so as to avoid thread serialization due to
branching as much as possible.  As for \gpuT,
Section~\ref{sec:temporal_indexing}, calculating $S$ on the host takes
negligible time.

As explained in the previous section, we enforce a minimum size for the
spatial subbins. Ensuring that subbins are not too small is necessary for
two reasons. First, with small subbins each entry segment could overlap
many subbins with high probability.  As a result, the query id would occur
many times in arrays $X$,$Y$, and/or $Z$, thereby wasting memory space on
the GPU and causing redundant calculations. Second, given our indexing
scheme and search algorithm described hereafter, a query that overlaps
multiple subbins along all three spatial dimensions may lead to duplicates
in the result set. These duplicates  would then need to be filtered out
(either on the GPU or the CPU).  To avoid duplicates, we simply default to
the purely temporal scheme whenever duplicates would occur. While this
behavior wastes computation (i.e., we lose spatial filtering capabilities),
the constraint on subbin size described in the previous section ensure that
it occurs with low probability.

The pseudo-code of the search algorithm is shown in
in Algorithm~\ref{alg:GPU_spatiotemporal}.
It takes the following arguments: 
(i)~the $X$, $Y$, and $Z$ arrays;
(ii)~the database ($D$);
(iii)~the query set ($Q$); 
(iv)~the schedule ($S$); 
(v)~the query distance ($d$);
and (vi)~the memory space to store the result set (\emph{resultSet}).
As in Algorithm~\ref{alg:GPU_temporal},
arguments that lead to array transfers between the host and the GPU
are shown in boldface.  
The algorithm begins by checking the global thread id and aborts if it is
greater than $|Q|$ (line~\ref{alglineST.abort}).
The query assigned
to the thread is acquired from $Q$
(line~\ref{alglineST.queryseg_into_private_memory}).
A helper array is constructed that holds pointers to the $X$, $Y$,
and $Z$ arrays (line~\ref{alglineST.selector}).
If schedule $S$ does not give a specification for one of the $X$, $Y$, or $Z$ arrays ($S$[gid].arrayXYZ = -1) then the
algorithm defaults to the temporal scheme
(line~\ref{alglineST.temporal_begin}).
Otherwise, 
the algorithm 
retrieves the pointer to the correct $X$, $Y$, or $Z$ array
(line~\ref{alglineST.XYZ})  and determines the index
range for the entry segments
(lines~\ref{alglineST.entryidmin_into_private_memory}-\ref{alglineST.entryidmax_into_private_memory}).
It then processes the entry segments
(line~\ref{alglineST.loop}) as Algorithm~\ref{alg:GPU_temporal}.

\begin{algorithm}
\caption{\gpuST kernel.}
\label{alg:GPU_spatiotemporal}
\begin{algorithmic}[1]

\begin{small}
\Procedure{\gpuspatiotemporal}{$X$,$Y$,$Z$,$D$,{$\mathbf{Q}$},{$\mathbf{S}$},$d$, {\bf resultSet}}

\State {gid} $\leftarrow$ getGlobalId()\label{alglineST.pre_start}
\LINEIF{{gid}$\geq${$|Q|$}}{\Return}\label{alglineST.abort}
\State {queryID} $\leftarrow$ gid \label{alglineST.queryseg_into_private_memory}

\State {arraySelector} $\leftarrow$ $\{X, Y, Z\}$\label{alglineST.selector}

\If {{$S$[gid].arrayXYZ $\neq$ -1}}\label{alglineST.spatiotemporal_begin}
  \State {arrayXYZ} $\leftarrow$ {arraySelector}[$S$[gid].arrayXYZ]\label{alglineST.XYZ}
  \State {entryMin} $\leftarrow$ $S$[gid].entryMin\label{alglineST.entryidmin_into_private_memory}
  \State {entryMax} $\leftarrow$ $S$[gid].entryMax\label{alglineST.entryidmax_into_private_memory}
  \ForAll {$i$ $\in$ $\{${entryMin}, $\ldots$, {entryMax}$\}$}\label{alglineST.loop}
 
  \State entryID = arrayXYZ[$i$]
  \State result $\leftarrow$ compare($D$[entryID],$Q$[queryID])\label{alglineST.compare}
  \If {result $\neq$ $\emptyset$}
  \State {\bf atomic:} {resultSet} $\leftarrow$ {resultSet} $\cup$ result\label{alglineST.found}
  \EndIf
  \EndFor \label{alglineST.spatiotemporal_end}
\Else \label{alglineST.temporal_begin}
  \State {Lines~\ref{algline.entryidmin_into_private_memory}-\ref{algline.endloop} in Algorithm~\ref{alg:GPU_temporal}.} 
\EndIf \label{alglineST.temporal_end}

\State \Return\label{alglineST.return}
\EndProcedure

\end{small}
\end{algorithmic}
\end{algorithm}

\section{Experimental Evaluation}\label{sec:exp_eval}

\subsection{Datasets}

To evaluate the performance of our various indexing methods we use 3
datasets (1 real world and 2 synthetic) of 4-dimensional trajectories (3
spatial + 1 temporal).  In previous work~\cite{Gowanlock2014c} we have
evaluated a purely temporal indexing scheme that shares the general
principles of the scheme described in Section~\ref{sec:temporal_indexing}
(but assuming that $Q$ cannot fit in GPU memory). In that work we evaluated
the performance of distance threshold search for datasets with varying
statistical temporal properties, and found the index to perform equally well
across these datasets.  In this work, based on our previous experience and
because we consider spatial and spatiotemporal indexing schemes, we use
trajectory datasets that vary in terms of their sizes and spatial properties (e.g., 
spatial trajectory density):
\begin{itemize}
\item \random: a small, sparse synthetic dataset;
\item \merger: a large, real-world astronomy dataset;
\item \dense:  a high density synthetic dataset that 
               is motivated by astronomy applications.
\end{itemize}

The \random dataset consists of 2,500 trajectories generated via random
walks over 400 timesteps, for a total of 997,500 entry segments. Trajectory
start times are sampled from a uniform distribution over the [0,100]
interval.

\begin{figure}[t]
\centering
        \subfigure[]{
            \includegraphics[width=0.30\textwidth]{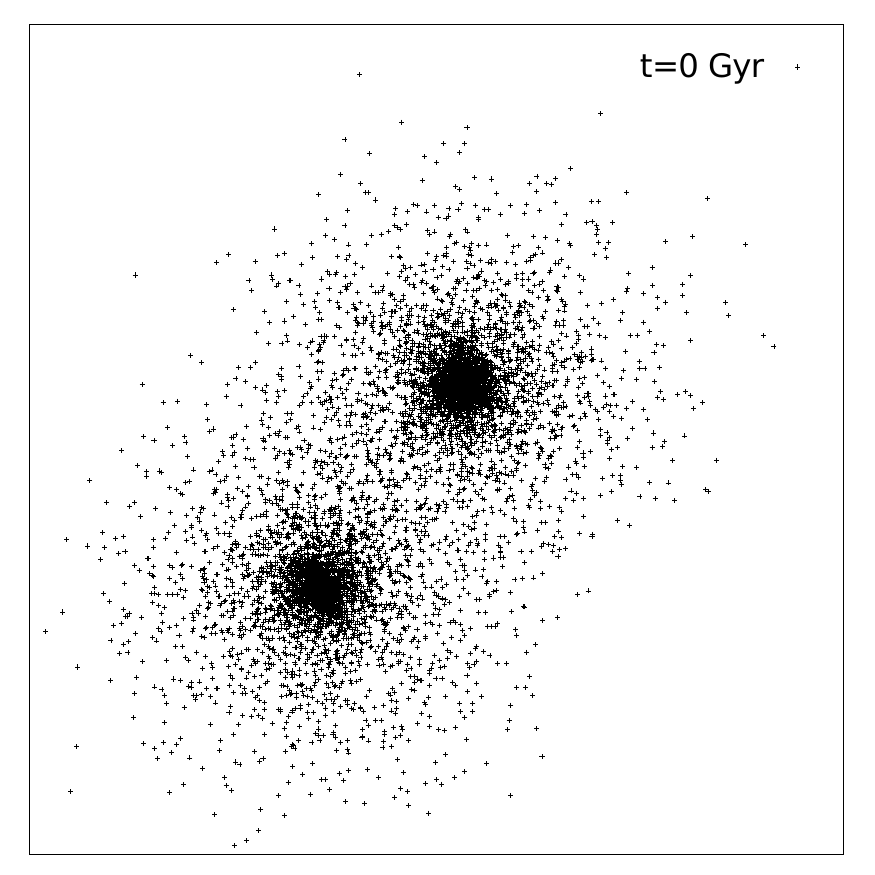}
  		}
        \subfigure[]{
            \includegraphics[width=0.30\textwidth]{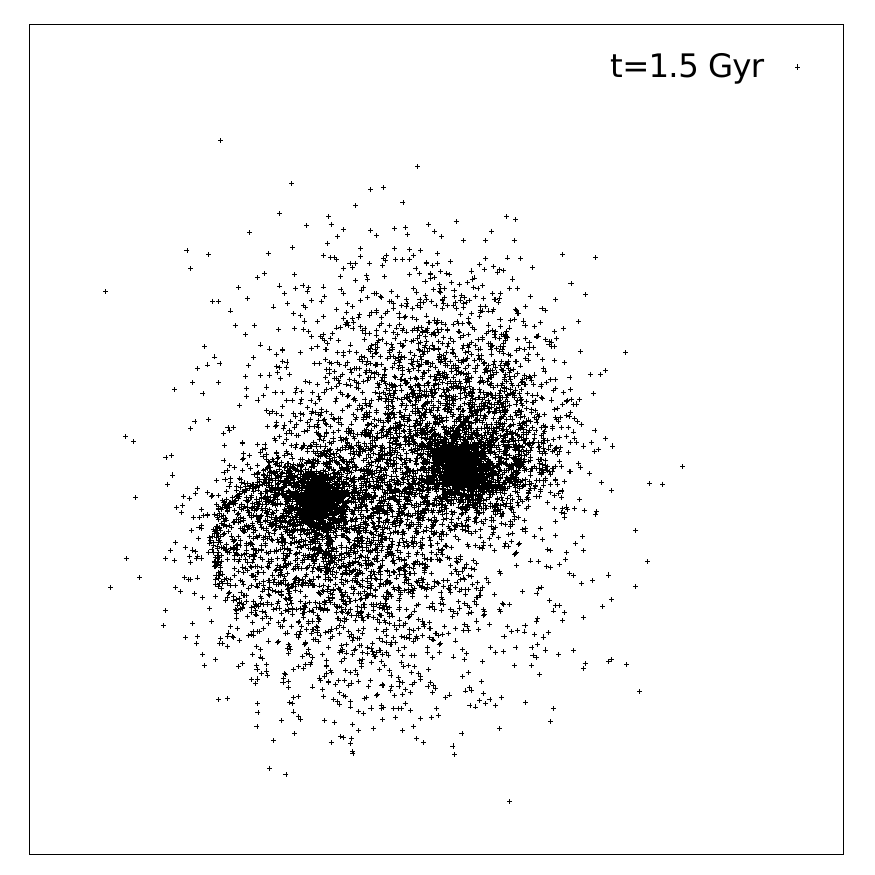}
        }
        \subfigure[]{
            \includegraphics[width=0.30\textwidth]{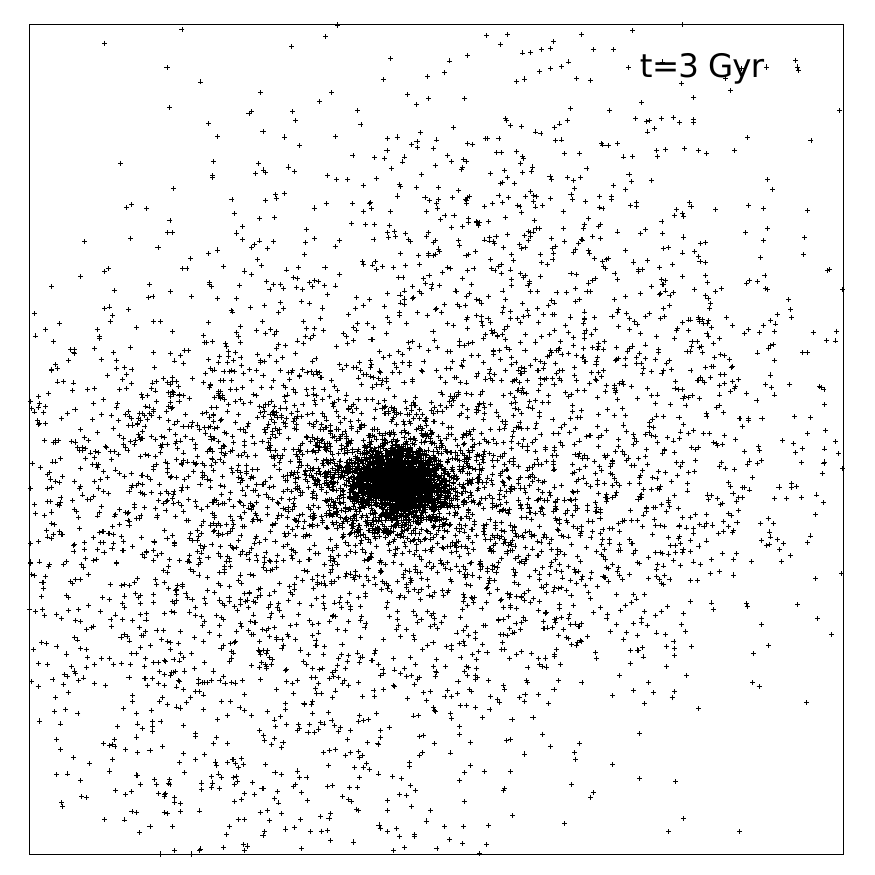}
        }
        
    \caption{Sample particle positions in the \merger dataset at times 0 Gyr (a), 1.5 Gyr (b) and 3 Gyr (c).}
   \label{fig:merger_dataset}
\end{figure}

The \merger dataset\footnote{This dataset was obtained from Josh
Barnes~\cite{1986Natur.324..446B}.} is from the field of astronomy and
consists of particle trajectories  that simulate the merger of the disks of
two galaxies.  It contains the positions of 131,072 particles over 193
timesteps for a total of 25,165,824 entry segments. Figure~\ref{fig:merger_dataset}
depicts particles positions projected onto the $x-y$ plane at different times, showing
the merger evolution.

%
%
%

%
%

The \dense dataset is generated as follows.  Consider the stellar number
density of the solar neighborhood, i.e., at galactocentric radius $R_\odot=8$ kpc
(kiloparsecs), of Reid et al.  \cite{2002AJ....124.2721R}, $n_\odot=0.112$
stars/pc$^{3}$. We develop a dataset with the same number of particles of
one disk in the \merger dataset (65,536), and 193 timesteps. To match the
density of \cite{2002AJ....124.2721R}, we require a volume of $65536/0.112
= 585142$ pc$^{3}$. This yields a cube with length, width and height
dimensions of 83.64 pc.  Note that we could have made the dataset more
spatially dense by picking a region close to the galactic center, since the
stellar density decreases as a function of $R$. We generate actual
trajectories as random walks as in the \random dataset, where all of the
particles are initially populated within the aforementioned cube.  We allow
the trajectories to move a variable distance in each of the x,y,z
dimensions at each timestep (between 0.001 and 0.005 kpc), and if a
particle moves outside of the cube by 20\% of the length of the cube in any
dimension, the particle is forced back towards the cube.   The particles,
on average, cannot travel too far from the cube such that we maintain a
fairly consistent trajectory density at each timestep.  This dataset aims to represent a density
consistent within the range of possible densities within the Milky Way that
a single node might process. The characteristics of each dataset are summarized in
Table~\ref{tab:datasets}.

\begin{table}
\centering
\caption{Characteristics of Datasets}
\begin{tabular}{|c|c|c|} \hline
Dataset&Trajec.&Entries\\ \hline
\hline
\random&2,500&997,500\\ 
\hline
\merger&131,072&25,165,824\\
\hline
\dense&65,536&12,582,912\\ 
\hline\end{tabular}
\label{tab:datasets}
\end{table}

\subsection{Experimental Methodology}

For all our distance threshold search implementations the GPU-side is
developed in OpenCL and the host-side is developed in C++. The host-side
implementation is executed on one of the 6 cores of a dedicated 3.46 GHz
Intel Xeon W3690 processor with 12 MiB L3 cache. The GPU-side
implementation runs on an Nvidia Tesla C2075 card with 6GiB of RAM and 448
cores. In all experiments we measure query response time as an average over
3 trials (standard deviation over the trials is negligible). We allocate a buffer
to hold the result set of the search on the GPU that can hold $5.0 \times 10^7$
items. In the description of the results we indicate when this buffer is
overcome, thus requiring incremental processing of the query. The response
time does not include the time to build the index or the time to store $D$
and the index in GPU memory. These operations can be performed off-line
before query processing begins.

We consider three experimental scenarios, each for one of our datasets:
\begin{itemize}
\item S1: The \random dataset and a query with 100 trajectories each with
400 timesteps for a total of 39,900 query segments.

\item S2: The \merger dataset and a query set with 265 trajectories each
with 193 timesteps for a total of 50,880 query segments.

\item S3: The \dense dataset and a query set with 265 trajectories each
with 193 timesteps for a total of 50,880 query segments.
\end{itemize}
For each scenario we use ranges of query distances (in units of kpc 
for S2 and S3).

In addition to our GPU implementations, we also evaluate a
CPU-only implementation.  This implementation relies on an in-memory R-tree
index, and is multithreaded using OpenMP.  Threads traverse the R-tree in
parallel, each for a different query segment, and return candidate entry
segments. This implementation was developed in our previous
work~\cite{Gowanlock2014,Gowanlock2014b}. In that work we investigated
``trajectory splitting," i.e., the impact of the number of segments stored
in each MBB in the R-tree index, $r$.  There is a trade-off between the
time to search the index (which decreases as $r$ increases due to lower
tree depth) and the time to process the candidate (which increases as $r$
increases due to higher index overlap).  All executions of the CPU
implementation  use 6 threads on our 6-core CPU. Results
in~\cite{Gowanlock2014c} show that this implementation achieves high
parallel efficiency.  Like for the GPU implementation, our response time
measurements do not include the time to build the index tree.

Although the experimental results in the following sections are constrained
by the specifics of our platform, the results of the CPU implementation are
used to demonstrate that the GPU can be used efficiently for distance
threshold searches.  A fundamental difference between the CPU
implementation and the GPU implementation is that the former relies on
index-tree traversal while the latter relies on flat indexing schemes. This
is because tree traversals on the GPU are problematic, e.g., due to thread
divergence slowdowns.

\subsection{Results for the \random Dataset}\label{sec:results_random}

In this section, we present results for the \random dataset, first giving
results for individual implementations and then combining results that make
it possible to compare the implementations.  The \random dataset is
representative of small and sparse datasets in which few or no entry
segments are expected to lie within distance $d$ of a query segment, i.e.,
with a low number of \emph{interactions}.

\begin{figure}[t]
\centering
  \includegraphics[width=0.5\textwidth]{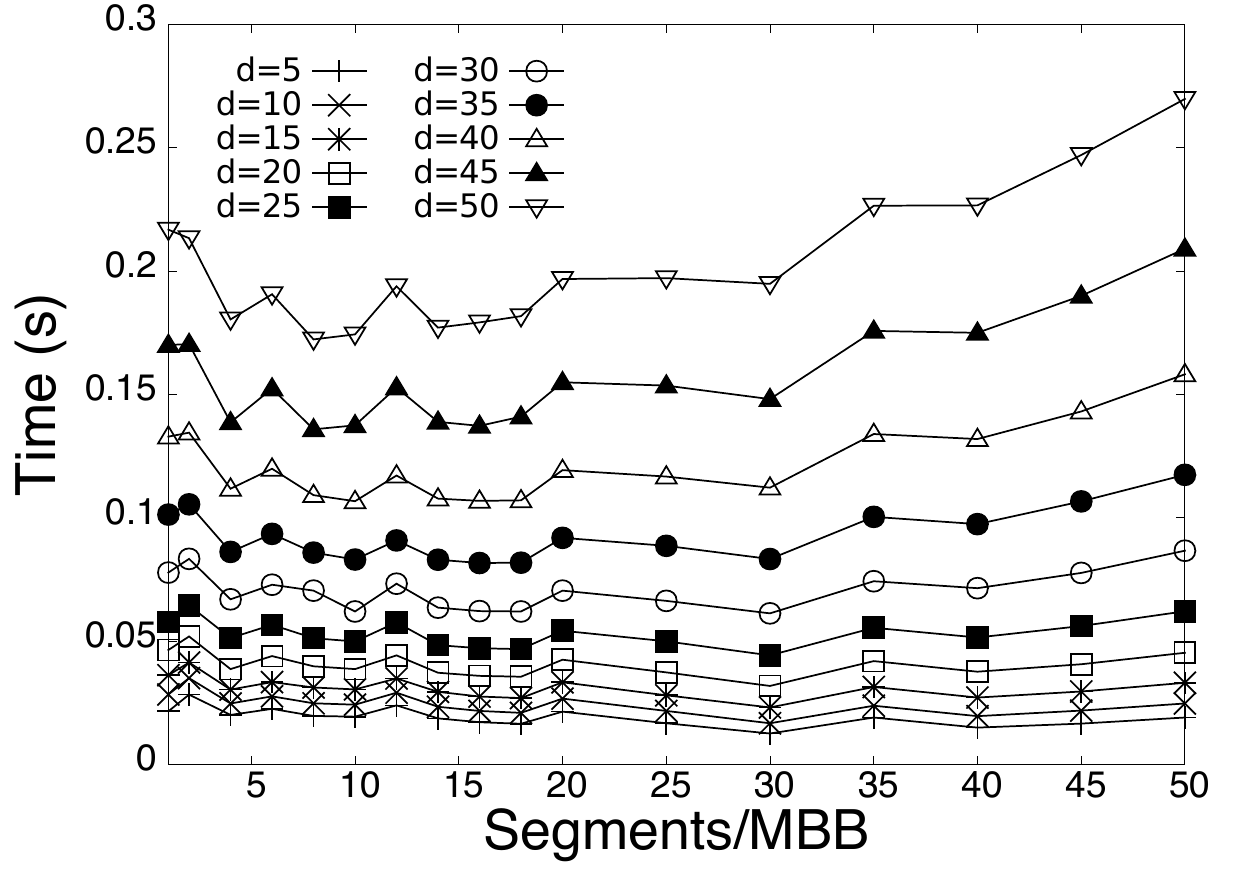}
    \caption{Response time vs. number of entry segments per MBB ($r$) for the CPU implementation in scenario S1 with $d=5, 10, \ldots, 50$.}
   \label{fig:CPU_random}
\end{figure}

Figure~\ref{fig:CPU_random} shows response time vs. the number of entry segments per MBB ($r$)
for the CPU implementation for a range of query distances. Using a single entry segment per MBB ($r=1$)
does not lead to the best response time. For this experimental scenario using, e.g.,
$r=10$ leads to good response times across all query distances.

\begin{figure}[t]
\centering
  \includegraphics[width=0.5\textwidth]{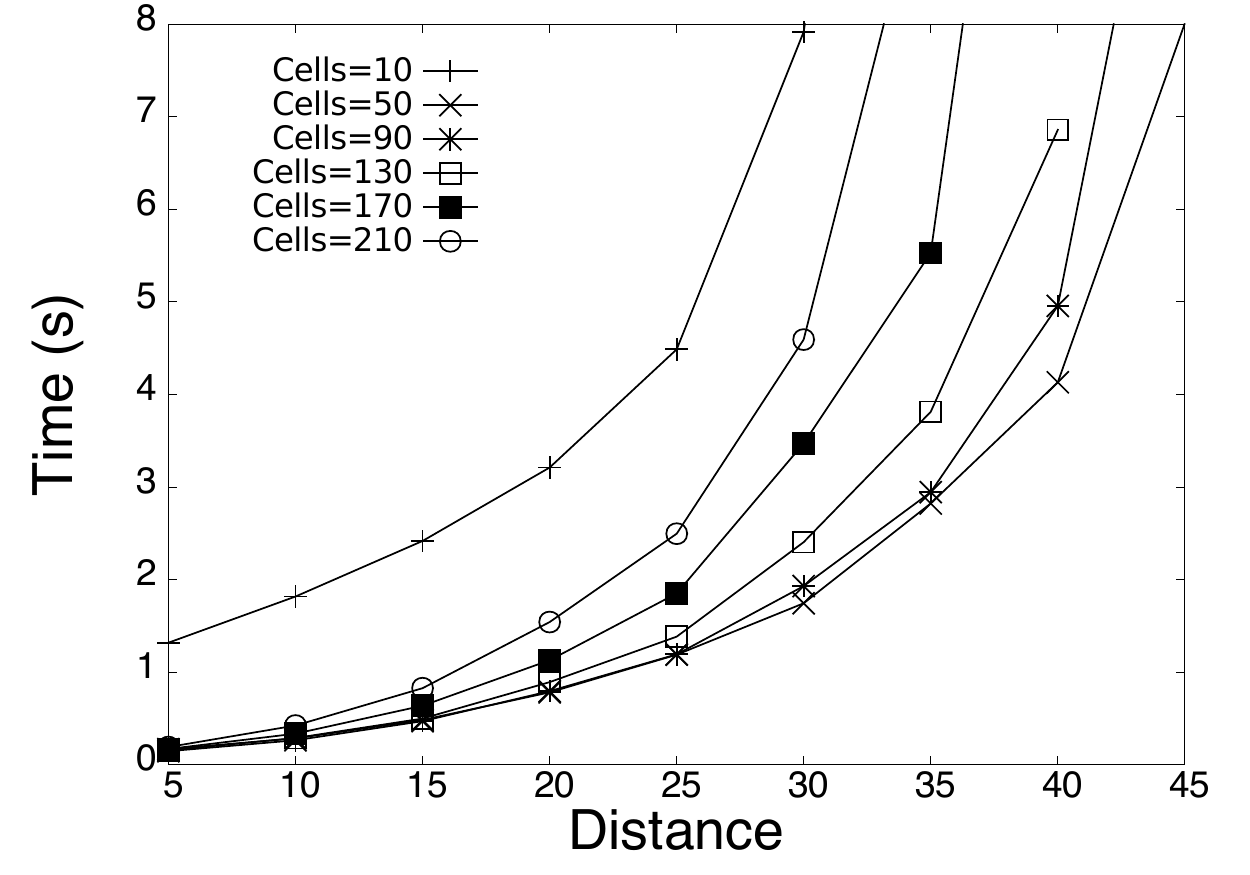}
    \caption{Response time vs. $d$ for \gpuS in scenario S1. Different curves are shown for different numbers of spatial cells in the $x$, $y$, and $z$ dimensions (i.e., ``Cells=10" means a $10\times 10\times 10$ grid).}
   \label{fig:GPU_FSG_random}
\end{figure}

Figure~\ref{fig:GPU_FSG_random} plots response time vs. $d$ for  \gpuS. Results are shown for a range of
grid sizes.  We use a total buffer size, $|U|$, of 2GiB to store
overlapping entry segments.  This is larger than the space necessary to
store $D$. This is thus an optimistic configuration for the FSG index.
Using too few grid cells leads, e.g., 10 per dimension, to poor performance
due to poor spatial selectivity. With poor spatial selectivity (i)~a large
candidate set must be processed and (ii)~many GPU threads overflow their
entry buffers $(U_k$) thus requiring multiple query processing attempts.
Likewise, using too many grid cells also leads to poor performance because
entry segments overlap multiple cells. As a result there is duplication of
index entries, and thus in the result set. Although filtering out these
duplicates takes negligible time, transferring them from the GPU back to 
the host incurs non-negligible overhead.  In
these experiments, and among the FSG configurations we have attempted,
using 50 cells per dimension leads to the best result.

Regardless of FSG configuration, we see rapid growth in response time as
$d$ increases.  The disposition of the FSG index to prefer small $d$ values
has also been alluded to in \cite{Zhang:2012:USH:2390226.2390229}. This
suggests that FSGs may not be particularly useful for spatiotemporal
trajectory searches due to the large spatial extent of the data and absence
of temporal discernment, unless query distances are small. However, a FSG
index is likely to perform well with fewer requirements, such as indexing
data with no temporal dimension, and/or focusing on point searches (instead
of line segments), which will not cause data duplication when a large
number of grid cells is used.

\begin{figure}[t]
\centering
  \includegraphics[width=0.5\textwidth]{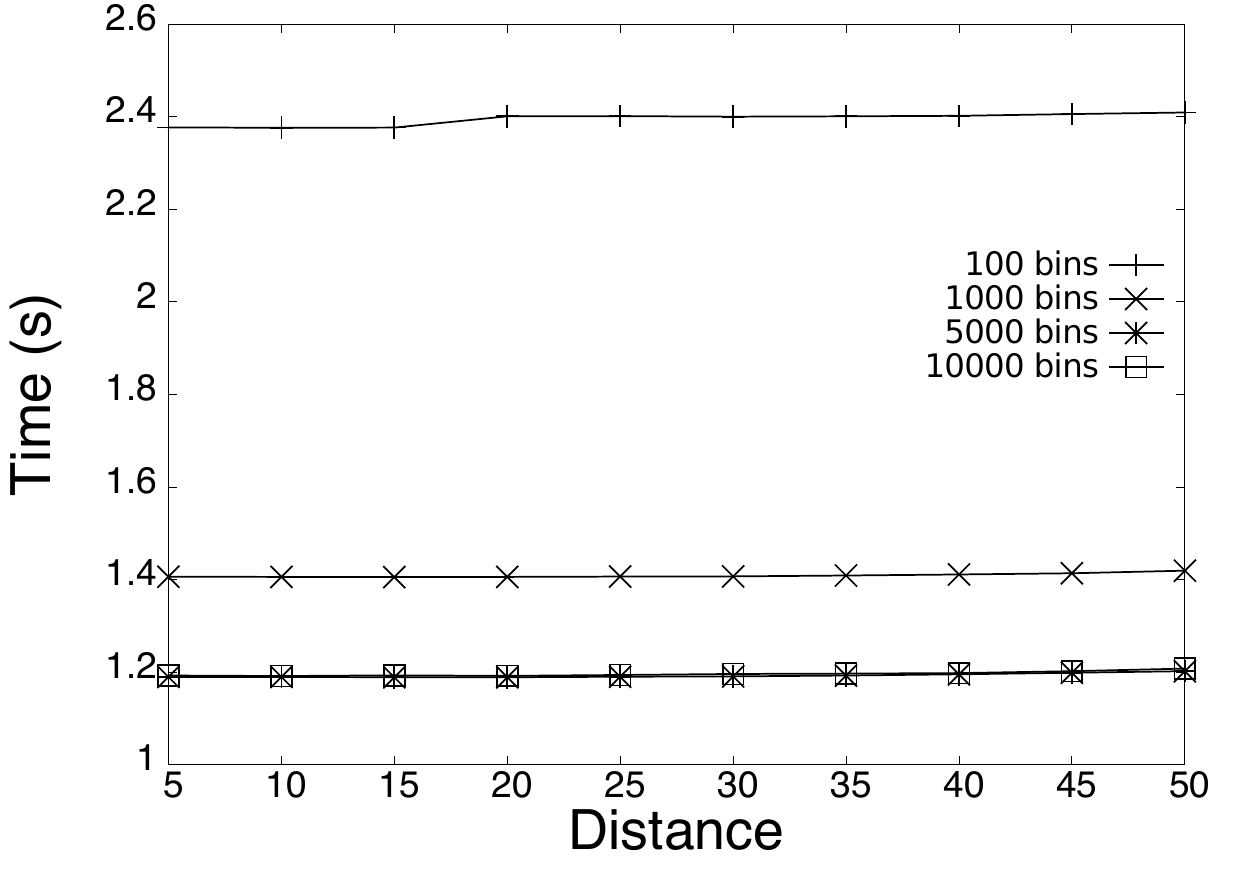}
    \caption{Response time vs. $d$ for \gpuT in scenario S1. Different curves are shown for different numbers of temporal bins (100, 1000, 5000, 10000).}
   \label{fig:GPU_temporal_random}
\end{figure}

Figure~\ref{fig:GPU_temporal_random} is shows response times vs. $d$
for \gpuT. Results are shown for a range of
number of temporal bins. Unlike for \gpuS, this
method is insensitive to the query distance.  With too few temporal bins
there is not enough temporal discrimination leading to large numbers of
interactions.  But as the number of bins increases the response time
reaches a minimum (increasing beyond 5,000 bins does not differentiate
entries as a function of temporal extent in the \random dataset).

\begin{figure}[t]
\centering
  \includegraphics[width=0.5\textwidth]{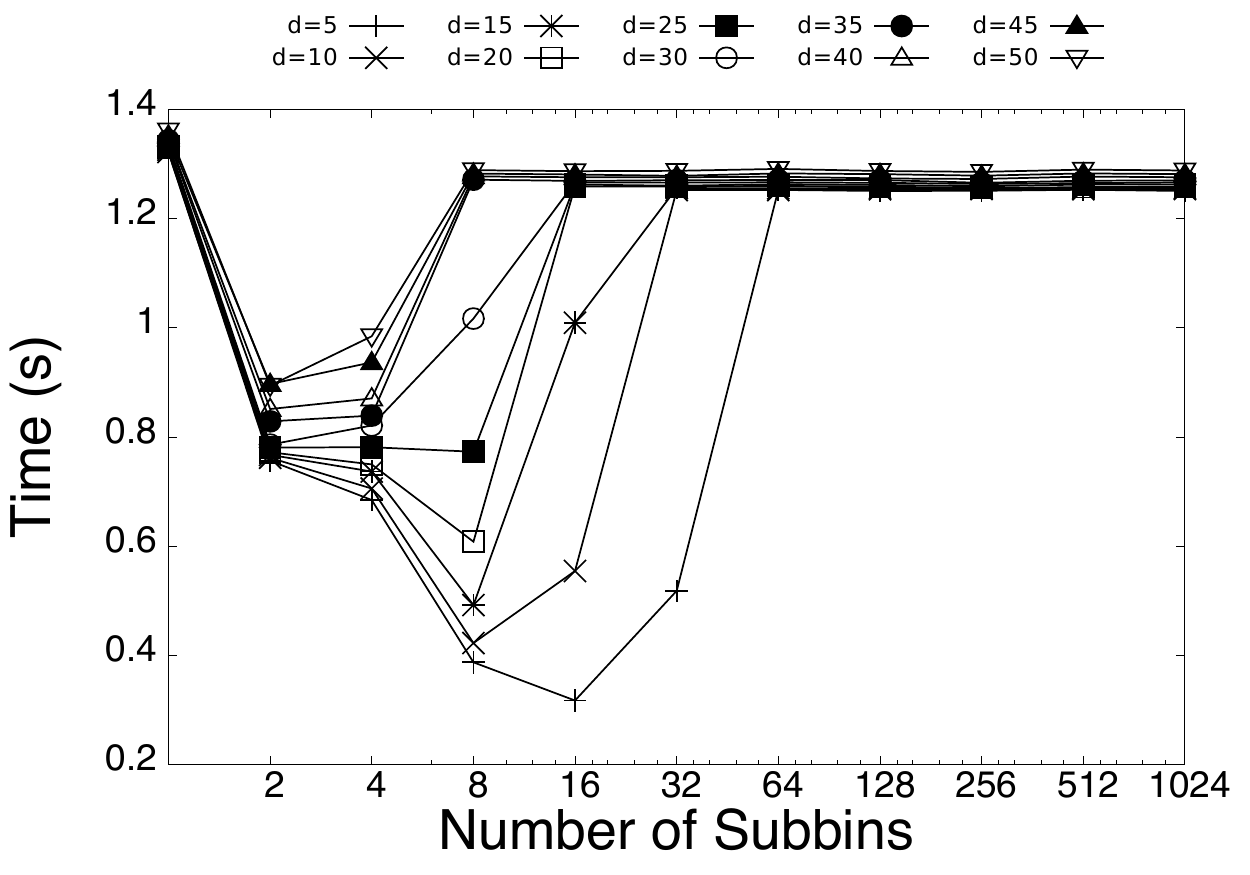}
    \caption{Response time vs. the number of subbins ($v$) for \gpuST in scenario S1. The number of temporal bins is set to 10,000. Different curves are shown for different query distances ($d=5,10,...,50$).} 
   \label{fig:GPU_spatiotemporal_random}
\end{figure}

Figure~\ref{fig:GPU_spatiotemporal_random} shows response time vs. the
number of subbins for \gpuST, where 10,000
temporal bins are used. Curves are shown for a range of $d$ values.  For
low $d$ values a greater number of spatial subbins is desirable. This is
because it is unlikely that a query will overlap multiple subbins, which
would cause our algorithm to revert to the purely temporal method, which
has no spatial selectivity.  As $d$ increases, queries overlap multiple
spatial subbins with higher probability. As a result, better performance is
achieved with fewer subbins.  Recall that we require that a query fall
within a single subbin so as to avoid duplication in the result set.
Without this requirement, an increasing number of subbins would suggest an
increase in the duplication of entries in the index, thereby increasing the
number of candidates that need to be processed (the same trade-off
discussed for \gpuS). There is thus a
trade-off between having too few or too many subbins, even when duplicates
in the result set are permitted.

We note that using 1 subbin in the spatiotemporal index is equivalent to
using a purely temporal index with no spatial selectivity.  Comparing
results between \gpuST with 1 subbin and \gpuT shows the effect of the additional indirection in the
spatiotemporal index. At $d=50$ (yielding the greatest number of
indirections in S1), with 1 subbin in the spatiotemporal index, the
response time is 1.36 s, whereas the response time is 1.21 s when using the
temporal index without any indirection. This is a 12.4\% increase in response time
due to the indirection. 

\begin{figure}[t]
\centering
  \includegraphics[width=0.5\textwidth]{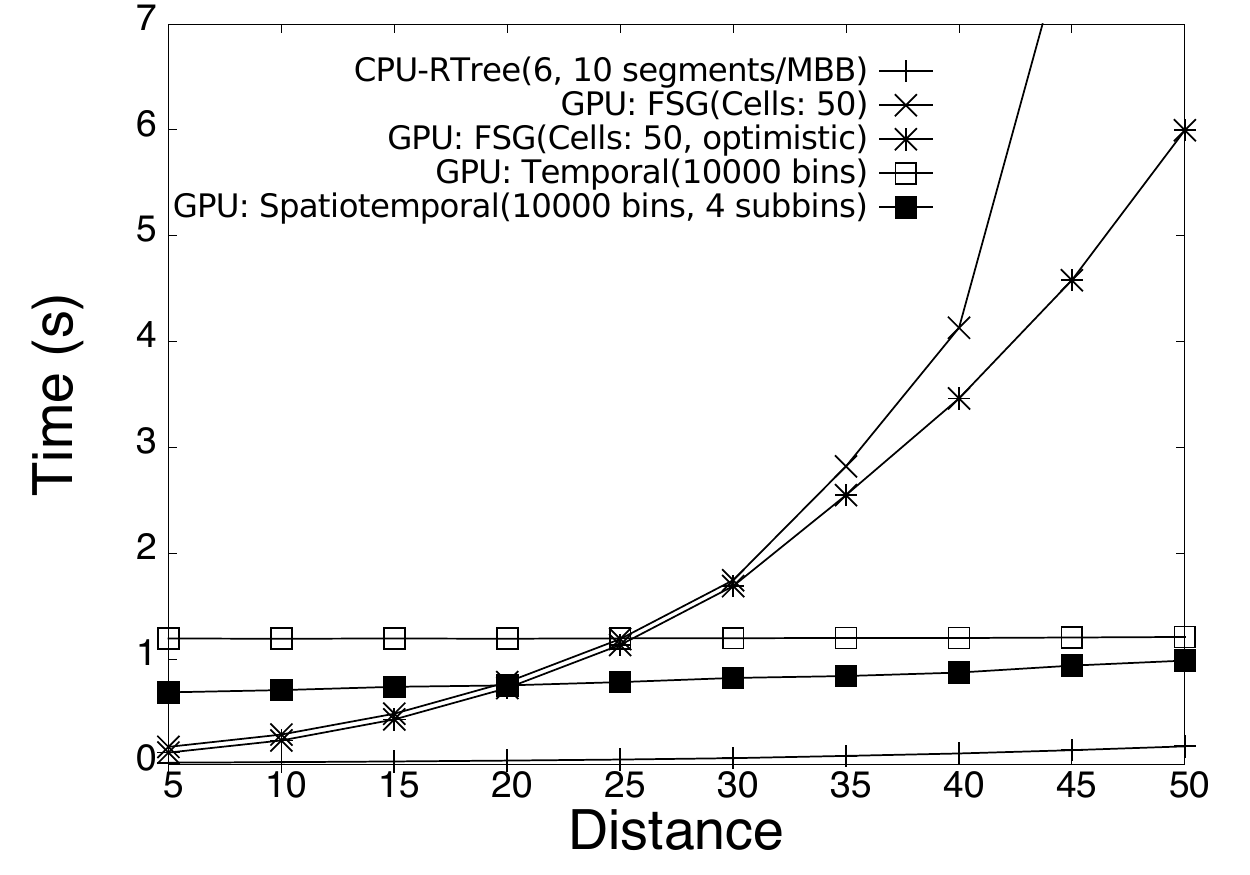}
    \caption{Response time vs. $d$ for our implementations for scenario S1.
    For the CPU implementation we use $r=10$ segments/MBB; for \gpuS we use 50 cells per spatial
    dimension; for \gpuT we use 10,000
    bins; and for \gpuST we use
    10,000 temporal bins and $v=4$ spatial subbins: For \gpuS we also plot an optimistic curve that
    ignores kernel re-launch overheads.}

   \label{fig:comparison_indexes_random}
\end{figure}

Figure~\ref{fig:comparison_indexes_random} shows response time vs. $d$ for
our four implementations. Each implementation is configured with best or
good parameter values based on previous results in this sections (see the
caption of Figure~\ref{fig:comparison_indexes_random} for details).  The
CPU implementation is best across all query distances.  
Comparing the GPU implementations, we see that \gpuS performs better than \gpuT and \gpuST when $d<20$, but that it does not scale well for larger
$d$ values. One may wonder whether this lack of scalability comes from the
overhead of re-launching the kernel due to buffer overflows.
Figure~\ref{fig:comparison_indexes_random} plots an ``optimistic" curve
that discounts this overhead. We see that the same trend, if not as
extreme, remains.  The temporal and spatiotemporal indexing methods have
consistent response times across query distances. Note that we could have
selected the best number of subbins for each value of $d$ from
Figure~\ref{fig:GPU_spatiotemporal_random}, which would have improved
results.  Comparing \gpuT and \gpuST, we see that
having spatial selectivity in addition to temporal indexing provides
performance gains, even on this small and sparse dataset.
We conclude that an in-memory R-tree is a good approach when indexing small
and sparse trajectory datasets that lead to few interactions. For such a
dataset, the overhead of using the GPU is simply too great.

\subsection{Results for the \merger Dataset}

In this section, we present results for our largest dataset, 
\merger, which contains over 25 million entry segments.  From Section~\ref{sec:results_random}, we find that the purely spatial FSG method leads
to extremely high response times for this larger dataset and as a result, we do
not consider it. 
In GPU executions, for some values of
$d$, we have to process $Q$ incrementally due to to insufficient space for
storing the full result set on the GPU. This is reflected in the measured
response times.

\begin{figure}[t]
\centering
  \includegraphics[width=0.5\textwidth]{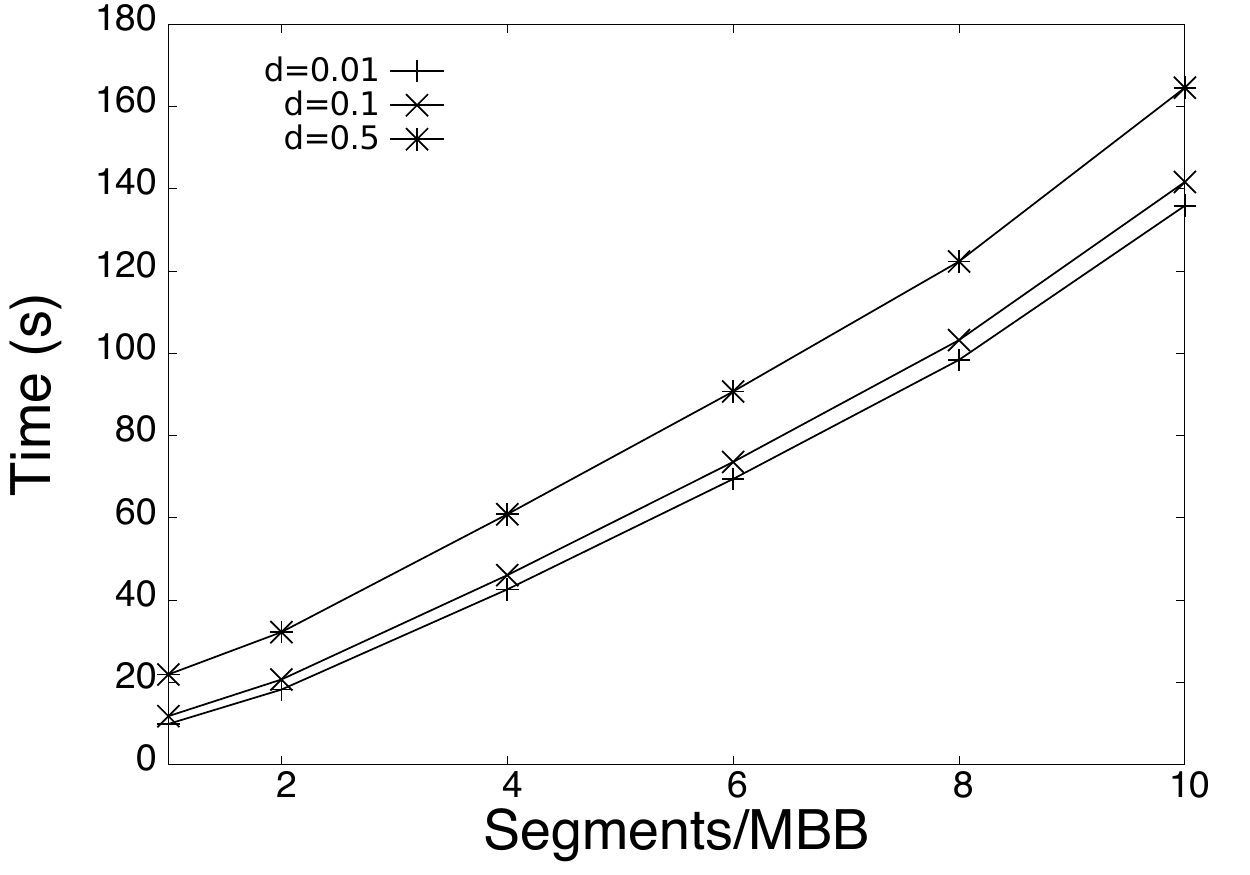}
    \caption{Response time vs. number of entry segments per MBB ($r$) for the CPU implementation in scenario S2 with $d=0.01, 0.1, 0.05$.}
   \label{fig:CPU_merger}
\end{figure}

Figure~\ref{fig:CPU_merger} shows response time vs. $r$ for the CPU
implementation for 3 query distances. With this large dataset, unlike with
\random, storing more than $r=1$ segments per MBB leads to higher response
time. A higher $r$ value decreases the time to search the R-tree index, but
this benefit is offset by the increase in candidate set size. This is an
important result. There is a literature devoted to assigning trajectory
segments to MBBs for improving response
time~\cite{Hadjieleftheriou:2002:EIS:645340.650233,Rasetic2005,Gowanlock2014}.
These works, however, do not consider large datasets.  For these datasets,
an intriguing future research direction is to take the opposite approach as
that advocated in the literature: splice individual polylines to increase
the size of the dataset (which can be thought of as setting $r<1$).


We do not show results for \gpuT as they are 
similar to those for the \random dataset. Using 1,000 temporal bins leads
to the lowest response time, which is consistent across all query distances. 

\begin{figure}[t]
\centering
  \includegraphics[width=0.5\textwidth]{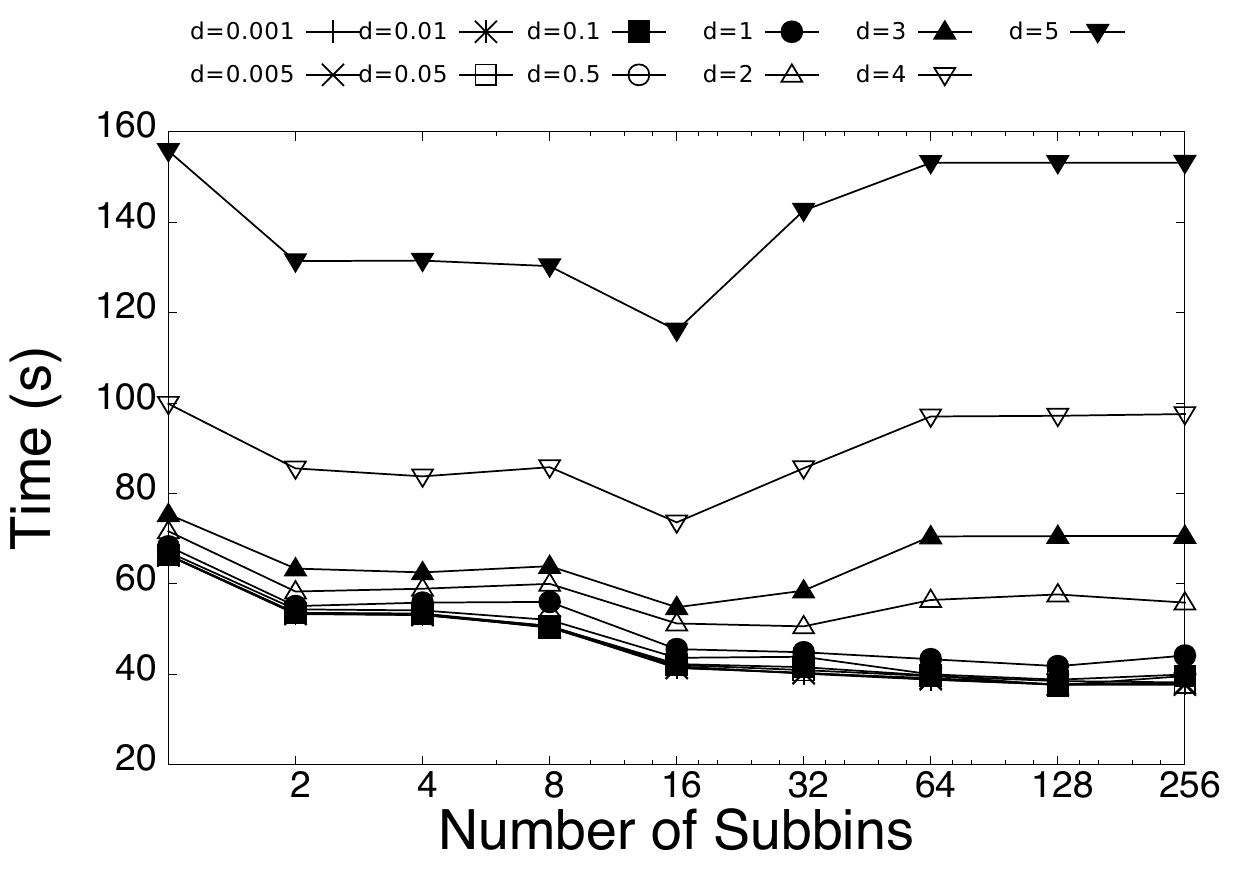}
    \caption{Response time vs. the number of subbins ($v$) for \gpuST in scenario S2. The number of temporal bins is set to 1,000. Different curves are shown for different query distances between $d=0.001$ and $d=5$.}
   \label{fig:GPU_spatiotemporal_merger}
\end{figure}


Figure~\ref{fig:GPU_spatiotemporal_merger} shows response time vs. number
of subbins for \gpuST, where  1,000 temporal bins are
used. Curves are shown for a range of $d$ values. A good number of subbins
is $v=16$ across all query distances, and this value is in fact best for
most query distances.  This implies that picking a good $v$ value can
likely be done for a dataset regardless of the queries.
Figure~\ref{fig:GPU_spatiotemporal_random} shows a dependency between $v$
and $d$ for the \random dataset. This dependency vanishes for a large
dataset with many interactions.



\begin{figure}[t]
\centering
  \includegraphics[width=0.5\textwidth]{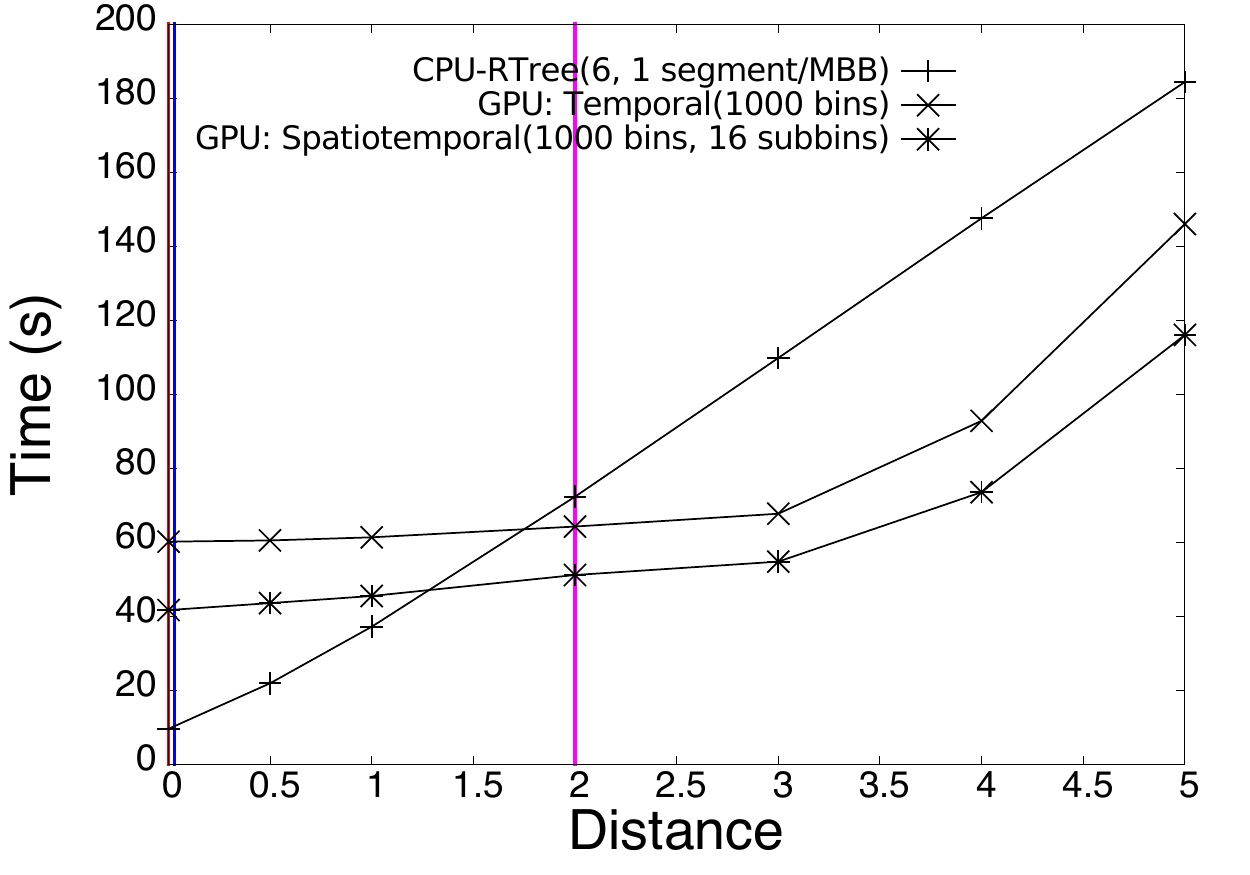}
    \caption{Response time vs. $d$ for our implementations for scenario S2.
	     For the CPU implementation we use $r=1$ segments/MBB; for \gpuT, we use 1,000 bins; for
	     \gpuST, we use 1,000
	     temporal bins and $v=16$ spatial subbins.
	     We indicate three distance thresholds that would
	     be interesting for the study of the habitability of the Milky
	     Way based on such datasets. Red: close encounters between stars and planetary systems
	     \cite{2013AsBio..13..491J}; Blue: supernova events on
	     habitable planetary systems \cite{2011AsBio..11..855G}, and
	     Magenta: studying the effects of gamma ray bursts on habitable
	     planets \cite{2005ApJ...634..509T}. Both the Red and Blue lines are
             close to the vertical axis.}
   \label{fig:merger_comparison_indexes}
\end{figure}

Figure~\ref{fig:merger_comparison_indexes} compares the performance of the
CPU implementation and \gpuT and \gpuST (\gpuS is omitted). Each method is configured with best or good parameter
values based on results in Figures~\ref{fig:CPU_merger}
and~\ref{fig:GPU_spatiotemporal_merger}. \gpuST
outperforms \gpuT across the board, with response
times at least 23.6\% faster. At low query distances the CPU implementation
yields the lowest response times. It is overtaken by \gpuST at $d\sim1.5$.  At $d=0.001$ the response time for the CPU
implementation is 9.70 s vs. 41.75 s for \gpuST (the GPU implementation is 330.4\% slower).  At $d=5$ these
response times become 184.4 s, and 116.09 s, respectively (the GPU
implementation is 58.8\% faster).  
We conclude that the GPU implementation outperforms the CPU implementation when using large datasets or when large query distances are considered.


\subsection{Results for the \dense Dataset}

We now present results for the \dense dataset, which is smaller than
\merger and representative of scenarios in which many trajectories are
located in a small spatial region, as motivated by the stellar number
density at the solar neighborhood. Note that increasing the density by even
$>4\times$ would still be consistent with that resembling the disk in the
inner Galaxy.

\begin{figure}[t]
\centering
  \includegraphics[width=0.5\textwidth]{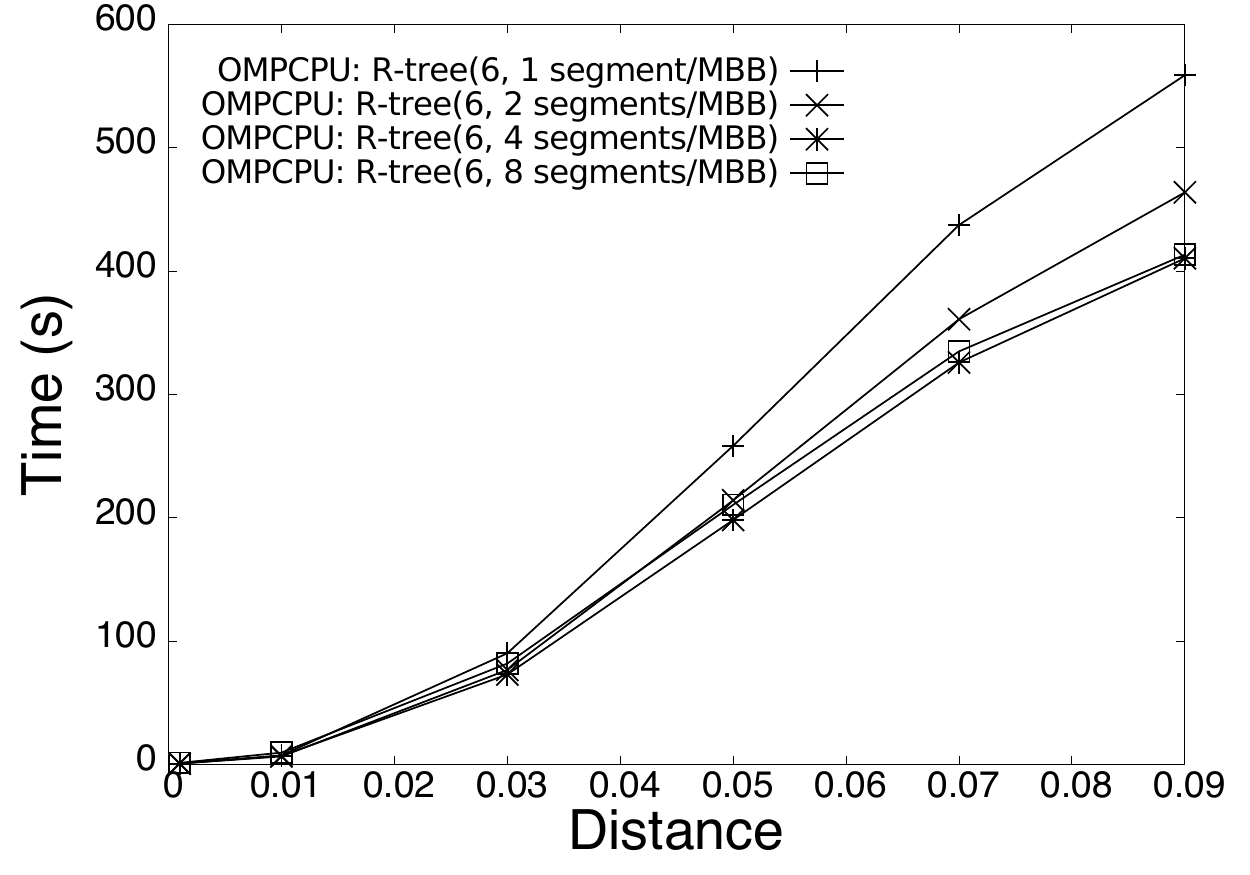}
    \caption{Response time vs. $d$ for the CPU implementation in scenario S3. Different curves
are shown for different values of $r$ (1,2,4, and 8).}
   \label{fig:RW_dense_CPU}
\end{figure}

Figure~\ref{fig:RW_dense_CPU} shows response time vs. query distance for
the CPU implementation for $r=1, 2, 4, 8$.
Unlike for \merger, which has $2\times$ the number of
entries as \dense, storing multiple segments/MBB improves response time. We
find that $r=4$ yields low response time values across all query
distances.  



As in the previous section, we do not show results for \gpuT
as they are similar to those for the \random dataset. Using
1,000 temporal bins leads to the lowest response time, which is consistent
across all query distances.

\begin{figure}[t]
\centering
        \subfigure[]{
            \includegraphics[width=0.45\textwidth]{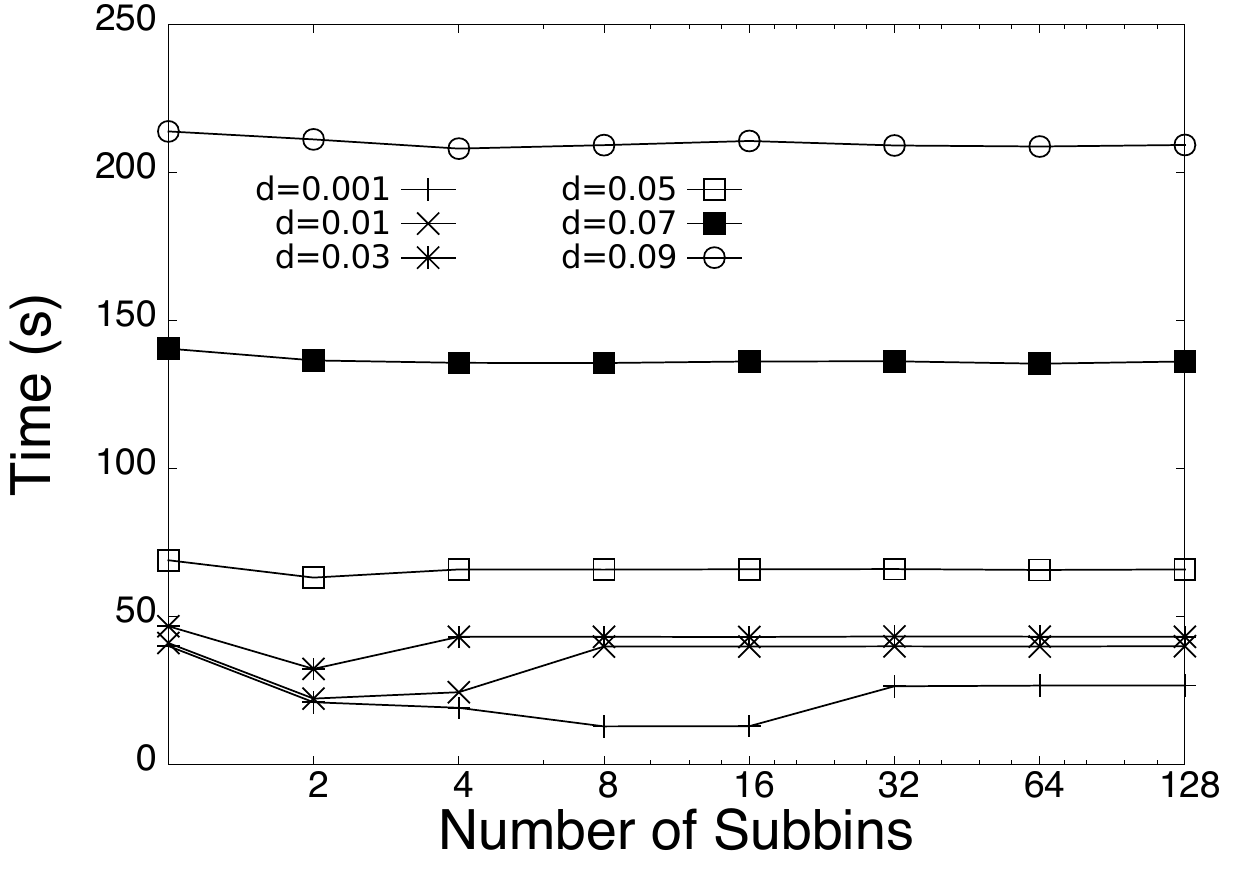}
  		}
        \subfigure[]{
            \includegraphics[width=0.45\textwidth]{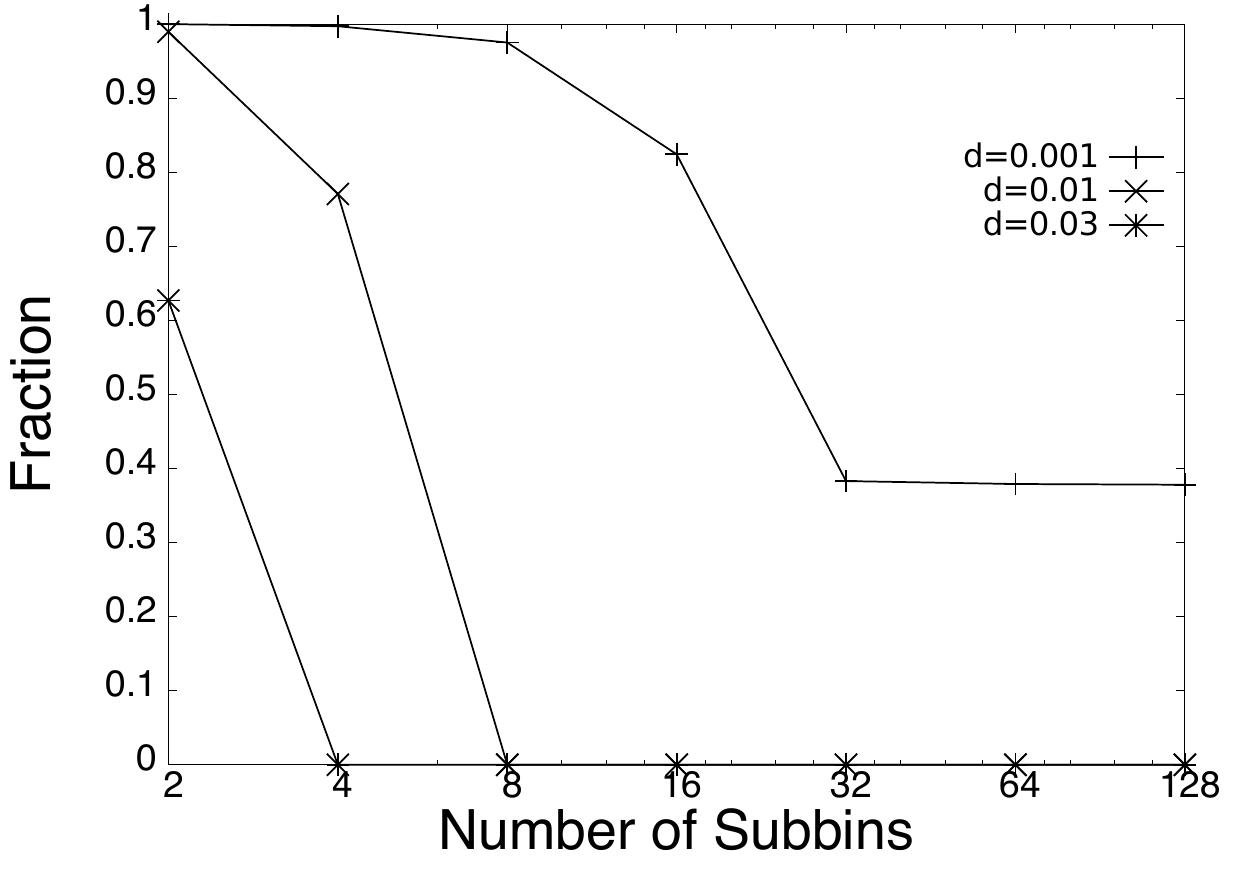}
        }
        
    \caption{(a)~Response time vs. number of subbins ($v$) for \gpuST for scenario S3 for a range of query distances. The number of temporal bins is set to 1,000. (b)~The fraction of queries that use the entries provided by subbins vs. the number of subbins ($v$).}
   \label{fig:RW_dense_spatiotemporal}
\end{figure}

Figure~\ref{fig:RW_dense_spatiotemporal}~(a) shows response time vs.
 the number of subbins ($v)$ for scenario S3 for \gpuST.  With this dataset, the use of subbins for reducing
response time is only possible for small query distances ($d=0.001, 0.01,
0.03$). This is because the dataset is smaller than \merger and because
with larger values of $d$, the queries are more likely to fall within
multiple subbins (in which case the search algorithm degenerates into a
purely temporal scheme).  Figure~\ref{fig:RW_dense_spatiotemporal}~(b)
shows the fraction of queries that utilized the entries provided by the
subbins for $d=0.001, 0.01, 0.03$.  Only the smallest query distance,
$d=0.001$, permits usage of the spatiotemporal index across a sizable
fraction of the number of subbins.  For instance for $d=0.03$ and $v=2$,
just over 60\% of the queries use the spatiotemporal index over the pure
temporal index, and when $v=4$, the entries provided by the spatiotemporal
index are not used. This explains why in
Figure~\ref{fig:RW_dense_spatiotemporal}~(a), there is no performance
improvement for $d>0.03$ when $v$ increases.

\begin{figure}[t]
\centering
  \includegraphics[width=0.5\textwidth]{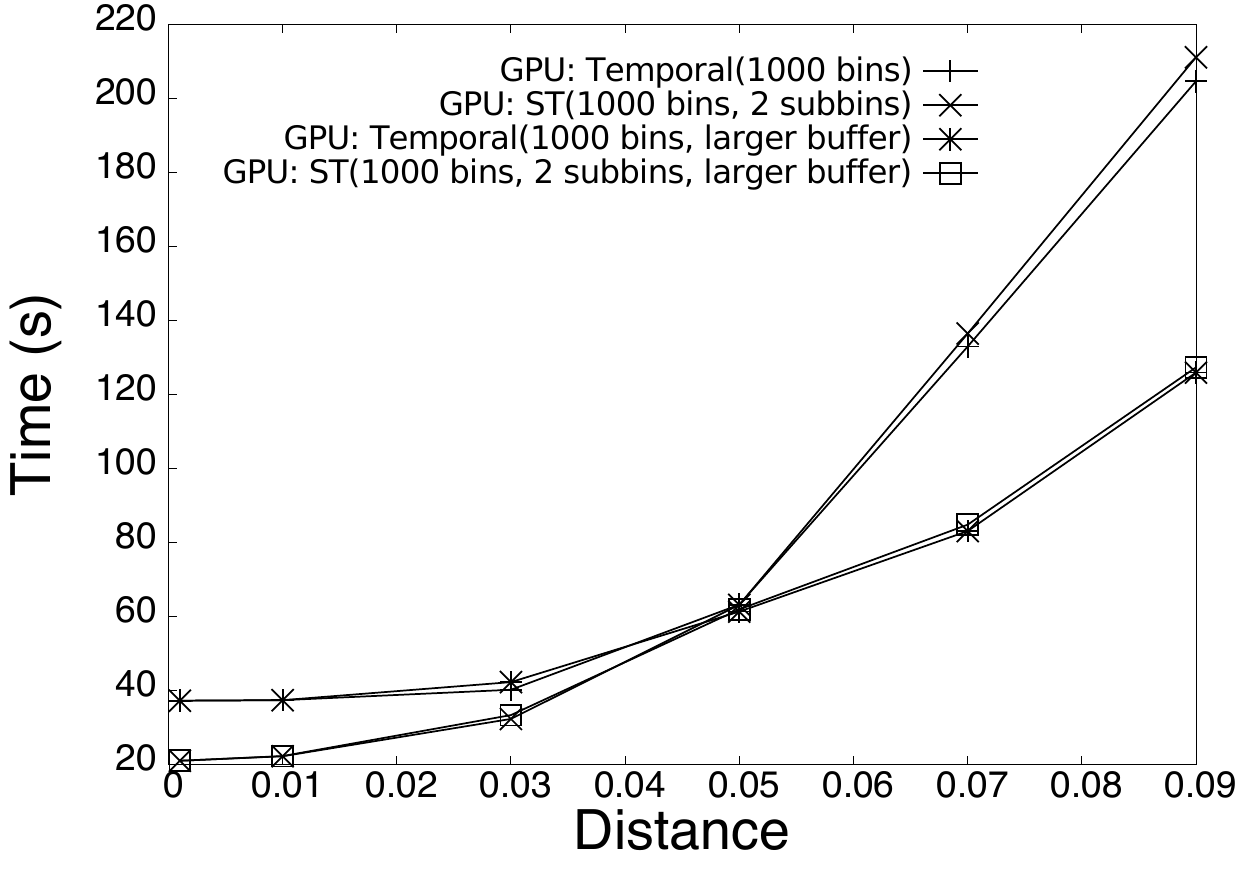}
    \caption{Response time vs. $d$ for \gpuT and \gpuST for scenario S3.  Results are shown for the original buffer size ($5\times10^7$) and for  a larger buffer size ($9.2\times10^7$).}
   \label{fig:RW_dense_comparison_buffer}
\end{figure}

Given the density of the dataset, for larger values of $d$, only a fraction
of the queries can be solved per kernel invocation as there is insufficient
memory space for the result set.  Since \dense has half as many entries as
\merger, we can increase the size of the buffer on the GPU for the
result set (from $5\times10^7$ elements for \merger to $9.2\times10^7$
elements for \dense).  Figure~\ref{fig:RW_dense_comparison_buffer} shows
the response time vs. $d$ for \gpuT and \gpuST with two buffer sizes.  Increasing the buffer
size by 84\% (thus requiring fewer kernel invocations) leads to decreases
in response time due to fewer host-GPU communications.  For instance, at
$d=0.09$ (which requires the greatest number of kernel invocations), the
spatiotemporal index, with $v=2$, using an increased buffer size for the
result set has a response time that is 65.76\% lower than with the initial
buffer size.  Although we could not run experiments with a larger buffer
size for scenario S2 (due to the large size of the \merger dataset), we
expect similar performance gains.  Since current trends point to
improvements in host-to-GPU bandwidth, in the future, our indexing
methods should provide even better performance improvements compared to CPU
implementations.


\begin{figure}[t]
\centering
  \includegraphics[width=0.5\textwidth]{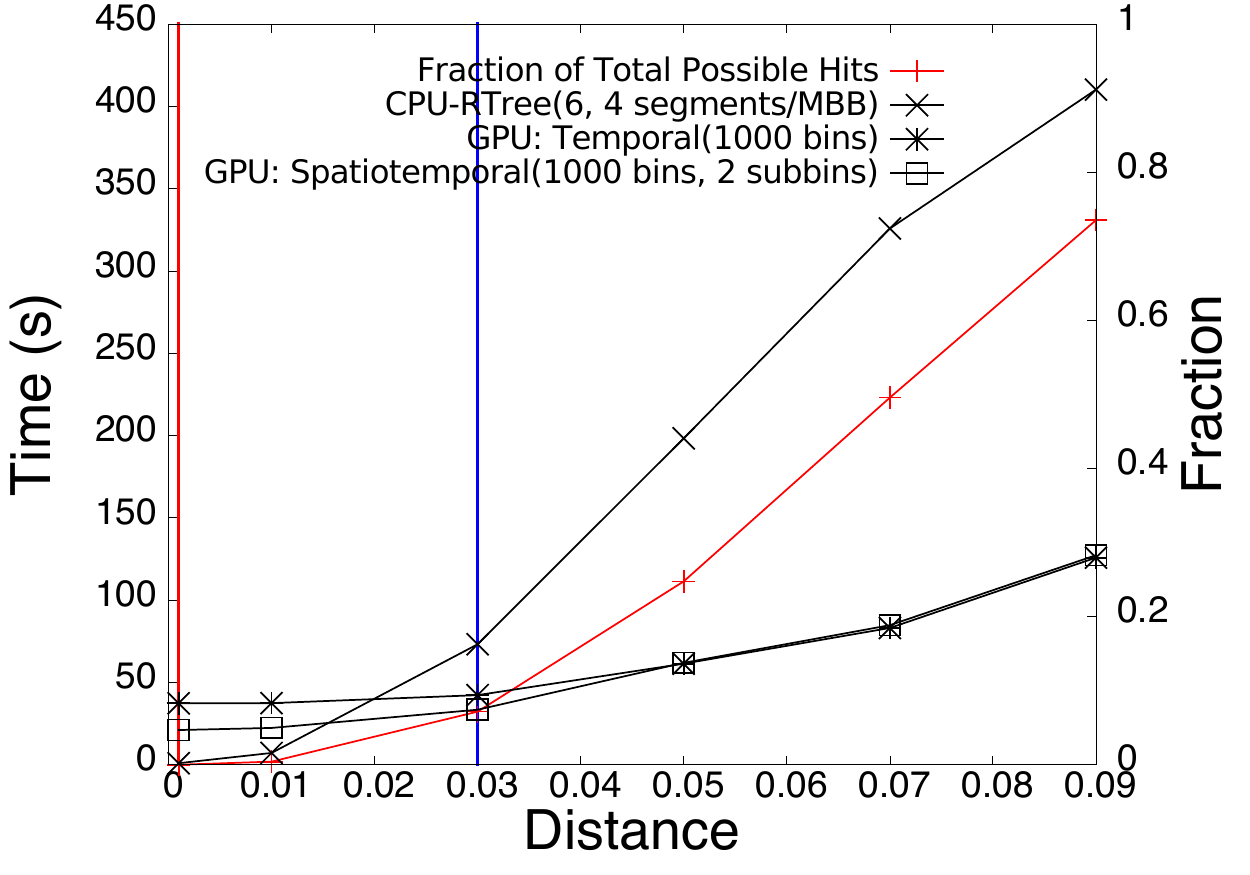}
    \caption{Response time (left vertical axis) and fraction of entries with distance $d$ of the query (right vertical axis) vs. $d$ for the CPU implementation, \gpuT, and \gpuST for scenario S3. 
For the CPU we show results for $r=1$ and $r=4$. 1,000 temporal bins are used for both the temporal and spatiotemporal indexing methods. $v=2$ spatial subbins are used for the spatiotemporal indexing method.}
   \label{fig:RW_dense_comparison_indexes}
\end{figure}

Figure~\ref{fig:RW_dense_comparison_indexes} shows response time vs. $d$
for the CPU implementation and \gpuT and \gpuST with the larger buffer sizes.  The query distance range
spans a wide range of result set sizes.  When $d=0.001$ $\approx 0\%$ of
the entries are within the query distance, and when $d=0.09$, 73.9\% of the
entries are within the query distance.  For very small query distances $d
\lesssim 0.02$, the CPU implementation yields the lowest response time, and
is outperformed by the GPU implementations for larger $d$.  For $d>0.03$,
\gpuST performs slightly worse than \gpuT.  This suggests that for dense datasets, when moderate to large query
distances are required, the pure temporal indexing method performs the
best.  At $d=0.05$, \gpuT is 223\% faster than
the CPU implementation (with $r=4$).

Comparing Figures~\ref{fig:merger_comparison_indexes} (\merger dataset)
and~\ref{fig:RW_dense_comparison_indexes} (\dense dataset), we see that the
range of query distances for which the GPU method is preferable to the CPU
method is much larger for the \dense dataset (considering the query
distances that correspond to relevant application scenarios -- the red,
blue, and magenta vertical lines). In the astronomy application domain, datasets
denser than the \dense dataset are relevant (i.e., to study 
the galactic regions at $R<8$ kpc). For these datasets a GPU approach will
provide even more performance improvement over a CPU implementation.



\begin{figure}[t]
\centering
        \subfigure[]{
            \includegraphics[width=0.45\textwidth]{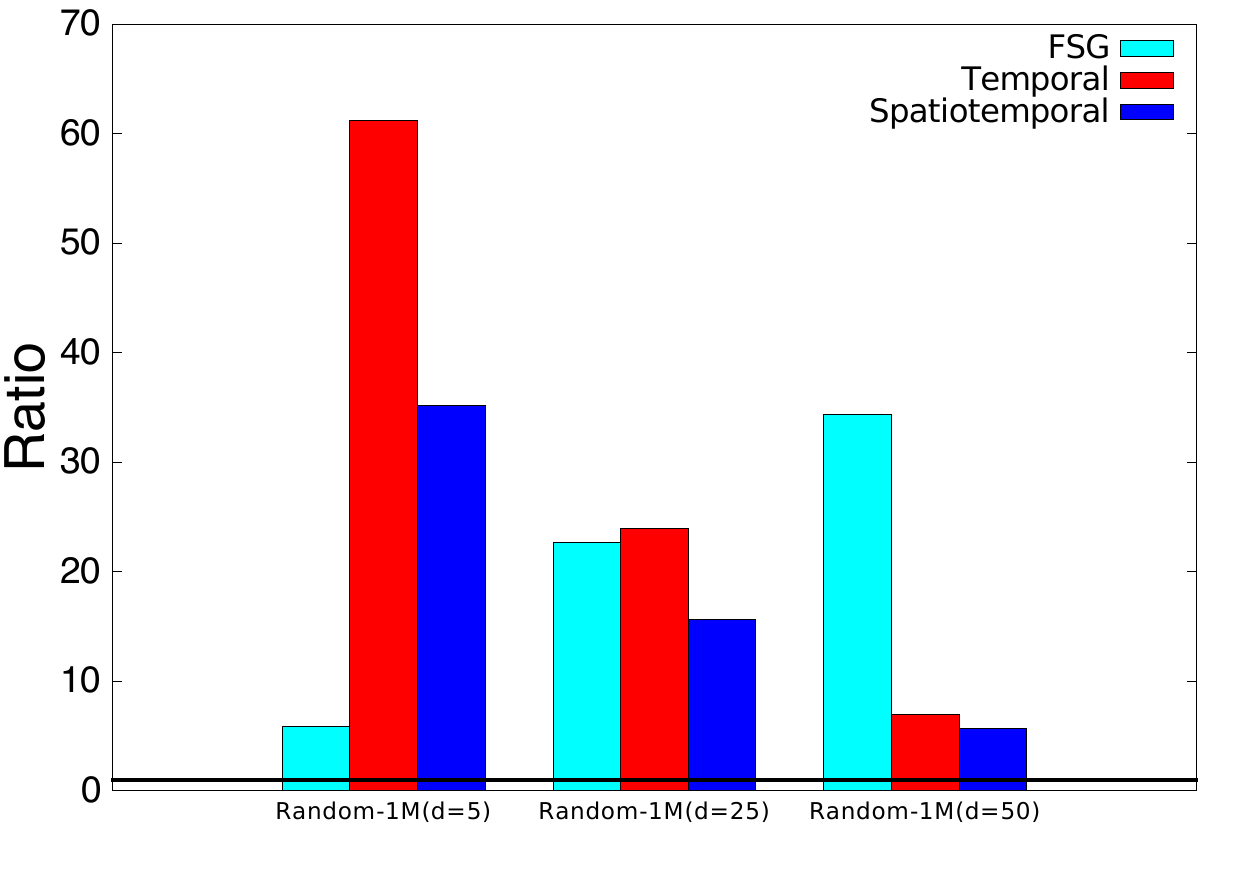}
  		}
        \subfigure[]{
            \includegraphics[width=0.45\textwidth]{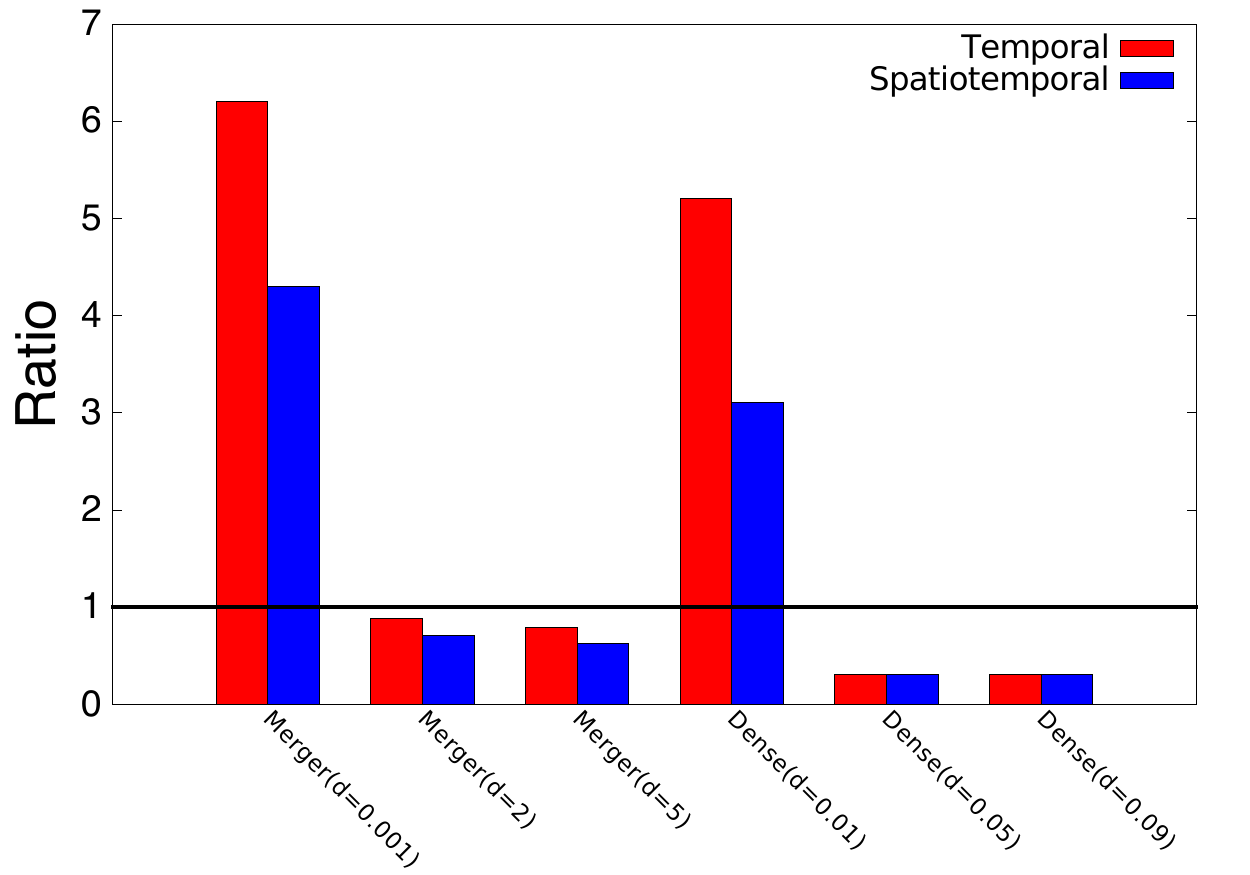}
        }
        
    \caption{Ratio of GPU to CPU response times across all datasets for (a) S1 and  (b) S2 and S3. Values below the $y=1$ line indicate improvements over the CPU implementation.}
   \label{fig:comparison_all_datasets}
\end{figure}

To summarize our results, Figure~\ref{fig:comparison_all_datasets} shows the
ratio of the response times of the GPU to CPU implementations for the 3
datasets for a few representative query distances. Data points below the
$y=1$ line correspond to instances in which the GPU implementation
outperforms the CPU implementation. The main findings are that
although the CPU is preferable for small and sparse datasets (Figure~\ref{fig:comparison_all_datasets}~(a)), the GPU
leads to significant improvements for large and/or dense datasets (Figure~\ref{fig:comparison_all_datasets}~(b)) unless query distances are very small.

\section{Conclusions}\label{sec:conclusions}

In this paper, we have proposed indexing methods and accompanying
algorithms for efficient distance threshold similarity searches on
spatiotemporal trajectory datasets.  Our main result is that GPU-friendly
indexing methods can outperform a multicore CPU implementation that uses an
in-memory R-tree index. This is the case when the datasets are large and/or
dense and the query distances are relatively large. Such scenarios are
routine in some applications, and in particular in our driving application
domain (astronomy). 
Overall, we find that spatiotemporal indexing
methods, which achieves selectivity both in time and space but without the
use of an index tree, is effective on the GPU.  
The trends and future plans for GPU technology point to
key improvements (faster host-to-GPU transfers, increased memory, etc.) that
will give a further advantage to GPU implementation of spatiotemporal
similarity searches. 

Our results show that for the in-memory R-tree CPU implementation, the
well-studied question of how to split a trajectory and store it in multiple
MBBs is not pertinent for large datasets. For these datasets, storing a
single segment by MBB is appropriate, and in fact it is likely appropriate
to splice segments and increase dataset size so as to trade-off higher
index-tree search time for small candidate sets to process.  This result
should apply to other similarity searches, such as $k$NN searches.

The main future work direction is to apply our indexing techniques to other
spatial/spatiotemporal trajectory searches and investigating hybrid
implementations of the distance threshold search that uses both the CPU and
GPU for query processing.

\section*{Acknowledgments}
The authors would like to thank Josh Barnes for providing us with the
\merger dataset.  This material is based upon work supported by the
National Aeronautics and Space Administration through the NASA Astrobiology
Institute under Cooperative Agreement No. NNA08DA77A issued through the
Office of Space Science.

\bibliographystyle{plain}

\end{document}